\documentclass[aps, prc, reprint, superscriptaddress]{revtex4-1}
\usepackage{amsmath}
\usepackage{amsfonts}
\usepackage{amssymb}

\usepackage{graphicx}
\usepackage[caption=false]{subfig}

\usepackage[colorlinks]{hyperref}
\hypersetup{linkcolor=blue, citecolor=blue, filecolor=blue, urlcolor=blue}
\usepackage[all]{hypcap}

\usepackage[utf8]{inputenc}

\usepackage{romannum}
\usepackage{booktabs}
\RequirePackage{lineno}

\usepackage{lipsum}

\newcommand{\muB}{$\mu_{\rm B}$}
\newcommand{\sNN}{$\sqrt {{s_{\rm NN}}}~$}

\begin{document}
\title{Effects of centrality fluctuation and deuteron formation on proton number cumulant in Au+Au collisions at \sNN = 3 GeV from JAM model}

\author{Arghya Chatterjee}
\author{Yu Zhang}
\author{Hui Liu}
\author{Ruiqin Wang}
\author{Shu He}
\affiliation{Key Laboratory of Quark \& Lepton Physics (MOE) and Institute of Particle Physics, Central China Normal University, Wuhan 430079, China}
\author{Xiaofeng Luo}
\email{xfluo@ccnu.edu.cn}
\affiliation{Key Laboratory of Quark \& Lepton Physics (MOE) and Institute of Particle Physics, Central China Normal University, Wuhan 430079, China}


\begin{abstract}
We studied the effects of centrality fluctuation and deuteron formation on the cumulants ($C_n$) and correlation functions ($\kappa_n$) of protons up to sixth order in most central ($b<3$ fm) Au+Au collisions at \sNN= 3 GeV from a microscopic transport model (JAM). The results are presented as a function of rapidity acceptance within transverse momentum $0.4<p_{T}<2 $ GeV/$c$. 
We compared the results obtained by centrality bin width correction (CBWC) using charged reference particle multiplicity with CBWC done using impact parameter bins. It was found that at low energies the centrality resolution for determining the collision centrality using charged particle multiplicities is not good enough to reduce the initial volume fluctuations effect for higher-order cumulant analysis. New methods need to be developed to classify events with high centrality resolution for heavy-ion collisions at low energies. Finally, we observed that the formation of deuteron will suppress the higher-order cumulants and correlation functions of protons and is found to be similar to the efficiency effect. This work can serve as a noncritical baseline for the QCD critical point search at the high baryon density region. 
\end{abstract}
\maketitle

\section{Introduction}

Quantum Chromodynamics (QCD) is the fundamental theory of the strong interaction. One of the major goals of high energy heavy-ion collision experiments is to explore the phase structure of the strongly interacting nuclear matter. The QCD phase structure can be displayed in the phase diagram, which is represented as a function of baryon chemical potential ($\mu_{B}$) and temperature ($T$)~\cite{Aggarwal:2010cw}. 
QCD based model calculations predict that at large \muB~the transition from hadronic matter to Quark-Gluon Plasma (QGP) is of first order. The end point of the first order phase transition boundary is known as QCD critical point (CP), after which there is no genuine phase transition but a smooth crossover from hadronic to quark-gluon degrees of freedom~\cite{Svetitsky:1985ye,Aoki:2006we,Gupta:2011wh}. Many efforts have been made to find the signature of the CP, theoretically~\cite{Fodor:2004nz,deForcrand:2002hgr,Qin:2010nq,Xin:2014ela,Shi:2014zpa,Fischer:2014ata,Lu:2015naa,Bazavov:2017tot,Fu:2019hdw,Fischer:2018sdj,Li:2018ygx,Yu:2018kvh} and experimentally~\cite{Aggarwal:2010wy, Adamczyk:2013dal, Luo:2015doi, Adamczyk:2014fia, Adamczyk:2017wsl}. However, the location of the CP and even the existence is still not confirmed yet~\cite{Rajagopal:1999cp}. The experimental confirmation of the QCD critical point would be a landmark in exploring the QCD phase structure.

In heavy-ion collisions, one of the foremost methods for the critical point search is through measurements of higher-order cumulants of conserved quantities, such as net-baryon, net-charge and net-strangeness number. Theoretically, it is expected that the higher-order cumulants of conserved charges are sensitive to the correlation length ($\xi$) of the system, which will diverge near the critical point~\cite{Stephanov:1998dy, Luo:2017faz,Bzdak:2019pkr,Luo:2020pef}. As a result non-monotonic variation of higher-order cumulant ratios from its baseline values are expected in existence of critical point. Furthermore,  theoretical calculation suggests that the ratio of sixth to second order cumulant ($C_6/C_2$) is sensitive to the phase transition and will become negative when the chemical freeze-out is close to the chiral phase transition boundary~\cite{Friman:2011pf,Bazavov:2020bjn}. Thus, the sixth order fluctuation could serve as a sensitive probe of the signature of the QCD phase transition~\cite{Nonaka:2020crv}. Experimentally, due to the detection inefficiency of neutral particles and multi-strange baryons, the net-proton and net-kaon are used as an experimental proxy of net-baryon and net-strangeness respectively. In the last few years, the measurement of second, third and fourth order cumulants of net-charge~\cite{Adamczyk:2014fia, Adare:2015aqk}, net-proton~\cite{Aggarwal:2010wy, Adamczyk:2013dal, Luo:2015doi}, and net-kaon~\cite{Adamczyk:2017wsl} multiplicity distributions have been conducted by the STAR and PHENIX experiment in the first phase of beam energy scan (BES-I, 2010-2017)  program at Relativistic Heavy Ion Collider (RHIC). The measurement of second order mixed cumulant have also been reported~\cite{Adam:2019xmk}. Recently, the HADES experiment published the proton number fluctuations in fixed target Au+Au collisions at \sNN= 2.4 GeV~\cite{Adamczewski-Musch:2020slf}.  Within current statistical uncertainties, the cumulants of net-charge and net-kaon distributions are found to have either modest or monotonic dependence on the beam energy, while the fourth order cumulant ratio ($C_4/C_2$) of the net-proton distributions exhibit non-monotonic behaviors as a function of \sNN, with a 3.1 $\sigma$ significance~\cite{Adam:2020unf}. To further confirm this non-monotonic behaviours, it is important to perform high precision fluctuation measurements at higher \muB~region. To fulfill this goal, RHIC has started the second phase of beam energy scan program (BES-II) since 2018, focusing on the collision energies below 27 GeV. From 2018 to 2020,  STAR experiment has taken the data of high statistics Au+Au collision at \sNN = 9.2, 11.5, 14.6, 19.6 and 27 GeV (collider mode) and  \sNN = 3.0 -- 7.7 GeV (fixed target mode). On the other hand, to understand various background contributions from different physics process, model (without CP) studies are important to provide baselines for the experimental search of the QCD critical point. These background contributions may arise from the limited detector acceptance/efficiency, initial volume fluctuation, autocorrelation and centrality resolution, centrality width, baryon number conservation and resonance decay. Some of those effects have been studied previously~\cite{Luo:2013bmi, Xu:2016qjd, Zhou:2017jfk,Chatterjee:2019fey, Chatterjee:2016mve, Ye:2018vbc, Zhang:2019lqz, Westfall:2014fwa, Zhou:2018fxx} and needs to be understood properly before making solid physics conclusions. 

In this paper, we studied the effects of centrality fluctuation and deuteron formation on the proton cumulant and correlation functions up to sixth order in most central Au+Au collisions at \sNN= 3 GeV using JAM model.
The paper is organized as follows. In section II, we briefly discuss the JAM model used for this analysis. In section III, we introduce the observables used for the present study. In section IV, we present the cumulants up to sixth order of proton multiplicity distribution at \sNN= 3GeV with JAM model and discuss the effect of centrality fluctuation and deuteron formation. The article is summarised in section V.

\section{The JAM model}
JAM (Jet AA Microscopic Transport Model) is a non-equilibrium microscopic transport model contracted on resonance and string degrees of freedom~\cite{Nara:1999dz, nara2019jam}. Hadrons and their excited states have explicit space and time propagation by the cascade method. Inelastic hadron-hadron collisions with resonance are applied at low energy whereas the string picture and hard parton-parton scattering are modeled at intermediate and high-energy respectively. The nuclear mean-field is applied based on the simplified version of the relativistic quantum molecular dynamics (RQMD) approach~\cite{Isse:2005nk}. Previously, JAM model has been used to compute several cumulants and studied different effects on particle number fluctuation in heavy-ion collision phenomenology~\cite{Zhang:2019lqz, he2016effects}. More details about JAM model can be found in reference~\cite{Nara:2016phs, nara2019jam, he2016effects}. In this study, we have analyzed around 25 million central events for Au+Au system at \sNN = 3 GeV generated using JAM model. Using the simulated events we calculated up to sixth order cumulants and correlation functions of event-by-event proton multiplicity distribution. The light nuclei like deuteron is not directly generated in JAM model, rather it is produced with an afterburner code along coalescence of nucleons with the phase space obtained from the JAM model~\cite{Liu:2019nii}. The coalescence conditions are constrained by relative distance ($\Delta R$) and relative momentum ($\Delta P$) in two body centre of mass frame. When the relative distance and momentum of any two nucleons are less than the given parameters ($R_{0},P_{0}$), the light nuclei are considered to be formed~\cite{Oh:2009gx,Sombun:2018yqh,Liu:2019nii,Deng:2020zxo}. Based on the charge rms radius of wave function for deuteron, we fixed the coalescence parameters of deuteron $\Delta R = 4$ fm and $\Delta P = 0.3$ GeV/$c$, respectively.

\section{Observables and methods} 
Higher-order multiplicity fluctuations can be characterized by different order cumulants ($C_{n}$). The $n^{th}$ order cumulant are expressed via generating function~\cite{Kitazawa:2017ljq} as,  
\begin{eqnarray}
C_{n} = \frac{\partial^{n}}{\partial \alpha^{n}} K(\alpha)|_{\alpha = 0},
\end{eqnarray}
where $K(\alpha)$ is the cumulant generating functions, which is logarithm of moment generating function ($K(\alpha) = \ln(M(\alpha))$). From event-by-event multiplicity distributions, the various order cumulants can be expressed in terms of central moment as follows: 
\begin{eqnarray}
C_{1} &=& \langle N \rangle, \\
C_{2} &=& \langle (\delta N)^{2} \rangle, \\
C_{3} &=& \langle (\delta N)^{3} \rangle, \\
C_{4} &=& \langle (\delta N)^{4} \rangle - 3\langle (\delta N)^{2} \rangle^{2}, \\
C_{5} &=& \langle (\delta N)^{5} \rangle - 10\langle (\delta N)^{2} \rangle \langle (\delta N)^{3} \rangle, \\
C_{6} &=& \langle (\delta N)^{6} \rangle - 15\langle (\delta N)^{4} \rangle \langle (\delta N)^{2} \rangle - 10\langle (\delta N)^{3} \rangle^{2} \\ \nonumber
&+& 30\langle (\delta N)^{2} \rangle^{3},
\end{eqnarray}  
where $N$ is the event-by-event particle number and $\delta N = N - \langle N \rangle$ represents the deviation of $N$ from its mean. $\langle ... \rangle$ represents an average over the event sample. The $n$-th order cumulant $C_{n}$ is connected to thermodynamic number susceptibilities of a system at thermal and chemical equilibrium.  
\begin{eqnarray}
C_{n} = VT^{3} \chi_{n},
\end{eqnarray}
where $V$ is the system volume, which is difficult to be measured in heavy-ion collisions. To cancel out the volume dependence different order cumulant ratios are measured as experimental observables which are related to the ratios of thermodynamic susceptibilities~\cite{Luo:2017faz, Cheng:2008zh}. 
\begin{eqnarray}
\frac{C_{2}}{C_{1}} = \frac{\chi_{2}}{\chi_{1}} = \frac{\sigma^{2}}{M},~\frac{C_{3}}{C_{2}} = \frac{\chi_{3}}{\chi_{2}} = S\sigma, \nonumber \\
~\frac{C_{4}}{C_{2}} = \frac{\chi_{4}}{\chi_{2}} = \kappa\sigma^{2},~\frac{C_{5}}{C_{1}} = \frac{\chi_{5}}{\chi_{1}},~\frac{C_{6}}{C_{2}} = \frac{\chi_{6}}{\chi_{2}}
\end{eqnarray}
where $M$, $\sigma$, $S$ and $\kappa$ are mean, sigma, skewness and kurtosis of the multiplicity distribution respectively. Besides, the multi-particle correlation function $\kappa_{n}$ (or factorial cumulant) can also be expressed in terms of single particle cumulants~\cite{Bzdak:2016sxg, Ling:2015yau}, 
\begin{eqnarray}
\kappa_{1} &=& C_{1}, \\ 
\kappa_{2} &=& -C_{1} + C_{2}, \\
\kappa_{3} &=& 2C_{1} - 3C_{2} + C_{3}, \\
\kappa_{4} &=& -6C_{1} + 11C_{2} - 6C_{3} + C_{4}, \\
\kappa_{5} &=& 24C_{1} - 50C_{2} + 35C_{3} - 10C_{4} + C_{5}, \\
\kappa_{6} &=& -120C_{1} + 274C_{2} - 225C_{3} + 85C_{4}  \\
&-& 15C_{5} + C_{6},  \nonumber
\end{eqnarray}
In case of Poisson distribution, higher-order correlation function $\kappa_{n} (n > 2)$ are equal to zero. Thus, $\kappa_{n}$ can be also used to quantify the deviations from the Poisson distributions.

In this study, we have analyzed around 25 million central events ($b < $ 3 fm) for Au+Au system generated using JAM model. Here we studied the effect of centrality fluctuation and deuteron formation up to sixth order cumulant of proton multiplicity distribution in different acceptance window.  

In heavy-ion collision experiment the collision centralities are usually defined by using charged particle multiplicities ($N_{ch}$) around mid-rapidity in which the smallest centrality bin is a single multiplicity value. 
To avoid autocorrelation effect, protons have been excluded from $N_{ch}$ within $|\eta| < 1$ for centrality selection. This centrality definition is called Refmult3~\cite{Adamczyk:2013dal}. 
For better statistical accuracy the cumulants values are reported in wider centrality bins. So the centrality bin width correction (CBWC) needs to be done in order to suppress volume fluctuations in a wide centrality bin~\cite{Luo:2013bmi}. Conventionally, the CBWC is done by calculating cumulants in each Refmult3 bin~\cite{Chatterjee:2019fey}. 
However, for such a low energy due to much less final state particle multiplicity, even a single multiplicity bin correspond to a wider initial volume fluctuation. We will discuss this effect later on by comparing the results of Refmult3-CBWC with impact parameter (b) CBWC. 
In CBWC techniques, as shown in Eq.~\ref{cbwc}, first the n$^{th}$ order cumulants ($C_n$) are calculated in each bin $i$ and then weight it by the number of events in each bin ($n_{i}$),
\begin{eqnarray}
C_n^{i} = \frac{\sum_{i}n_{i}C_n^{i}}{\sum_{i}n_{i}}, 
\label{cbwc}
\end{eqnarray}
where $C_n^{i}$ is the n$^{th}$ order cumulant in $i$-th bin (either in b=0.1 fm bin or in each Refmult3 bin) and ($\sum_{i}n_{i}$) represents the total number of events. The uncertainties reported in the results are statistical due to finite size of event sample and obtained using standard error propagation method, called Delta theorem~\cite{kendall1963advanced, Luo:2011tp, Luo:2014rea}. Generally the uncertainty on cumulant measurement is inversely proportional to the number of events and proportional to the certain power of the width of the proton multiplicity distributions. 

\begin{figure}[htp!]
	\centering 
	\hspace{-1.cm} 
	\includegraphics[width=0.5\textwidth]{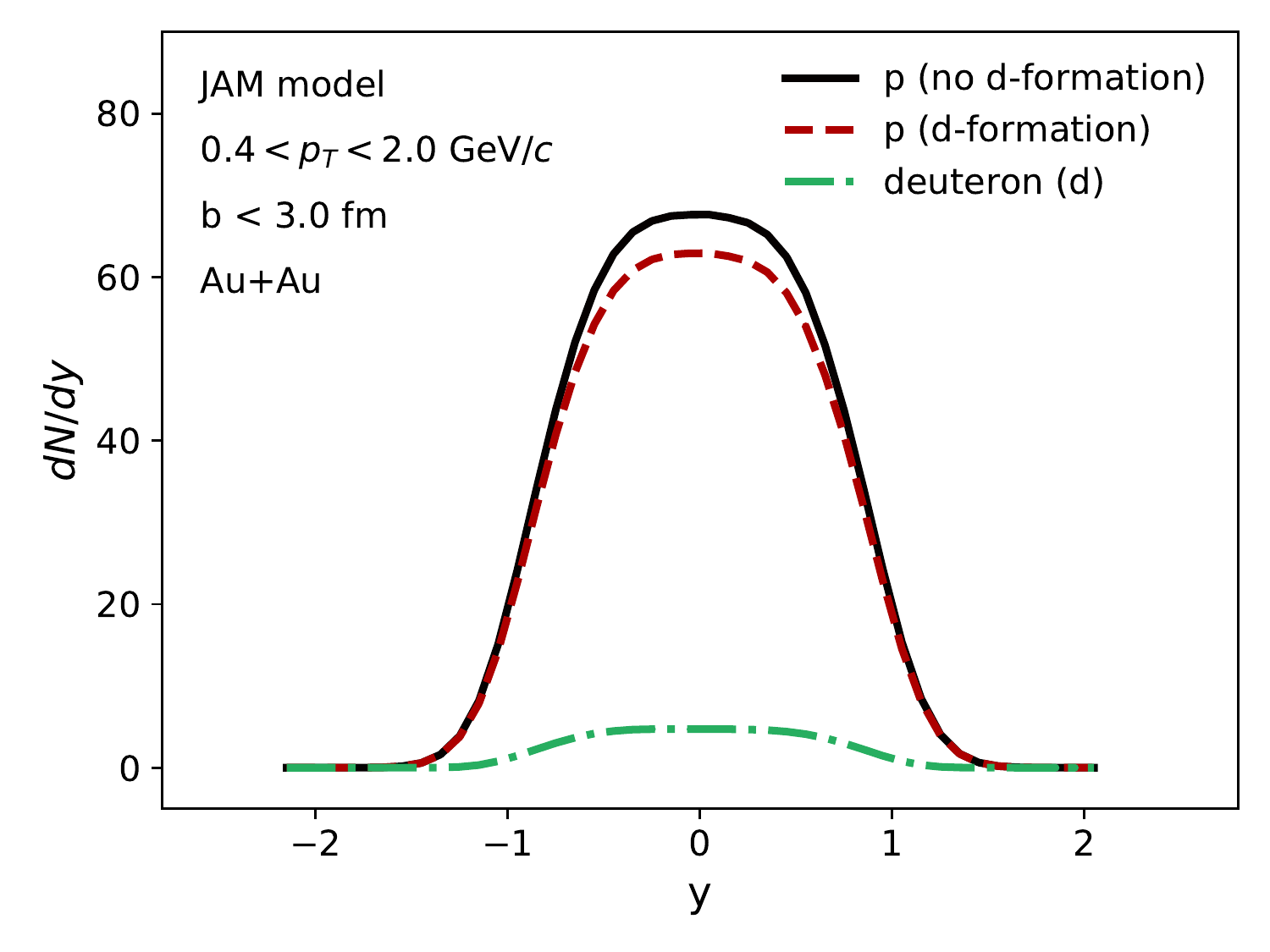}
	\caption{(Color online) Rapidity ($dN/dy$) distributions for proton with and without deuteron formation in most central ($b < 3$ fm) Au+Au system at \sNN = 3 GeV in JAM model.}
	\label{rapidity}
\end{figure}
\begin{figure}[htp!]
	\centering 
\hspace{-1.cm} 	
	\includegraphics[width=0.5\textwidth]{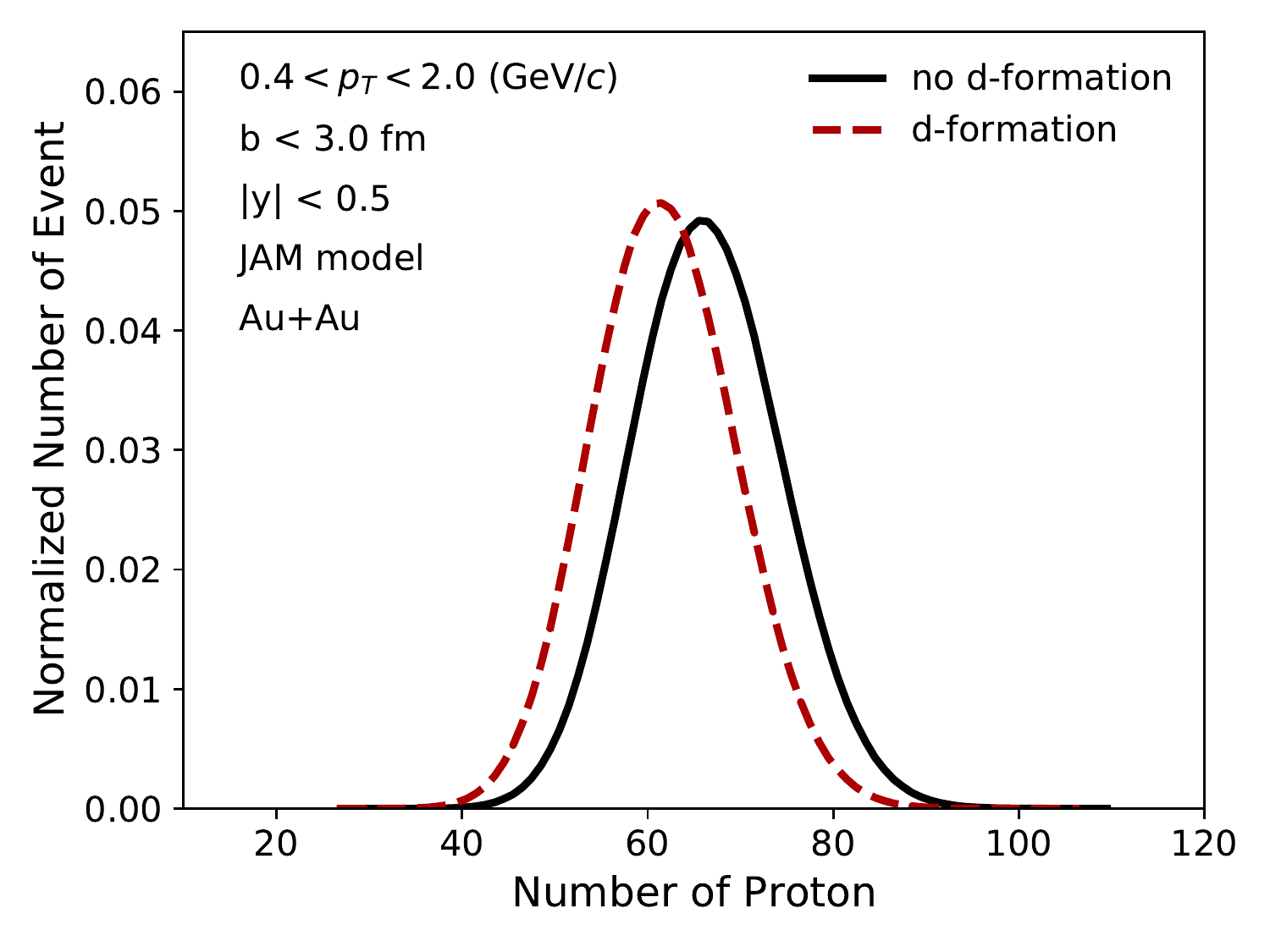}
	\caption{(Color online) Normalized event-by-event proton multiplicity distributions in most central ($b < 3$ fm) Au+Au collision at \sNN = 3 GeV with and without deuteron formation in JAM model.}
	\label{netP}
\end{figure}

\begin{figure}[htp!]
	\centering
	\subfloat{ 
		\label{fig:y equals x}
		\includegraphics[width=0.5\textwidth]{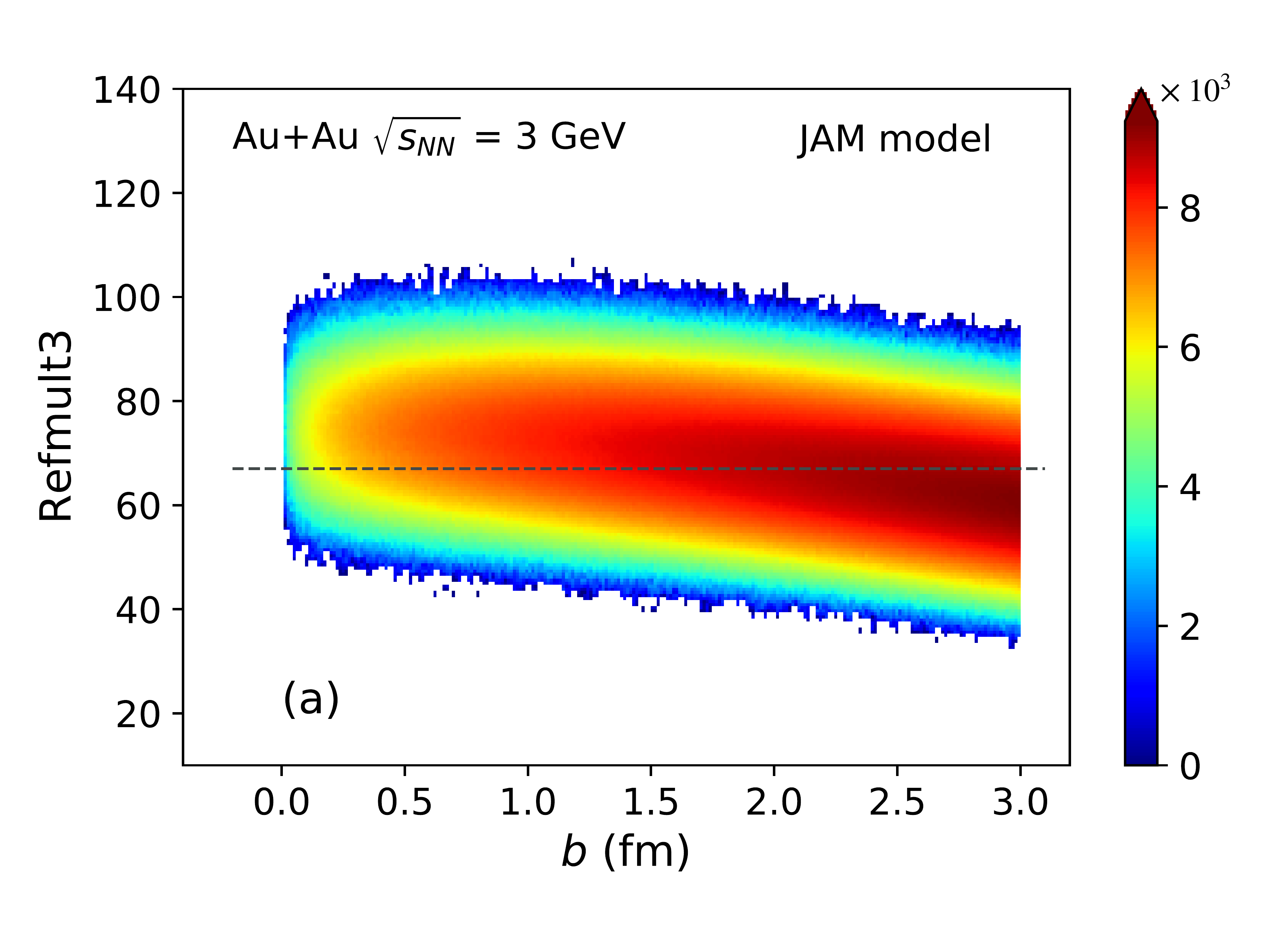}}
	\hfill
	\subfloat{
		\label{fig:five over x}
		\includegraphics[width=0.45\textwidth]{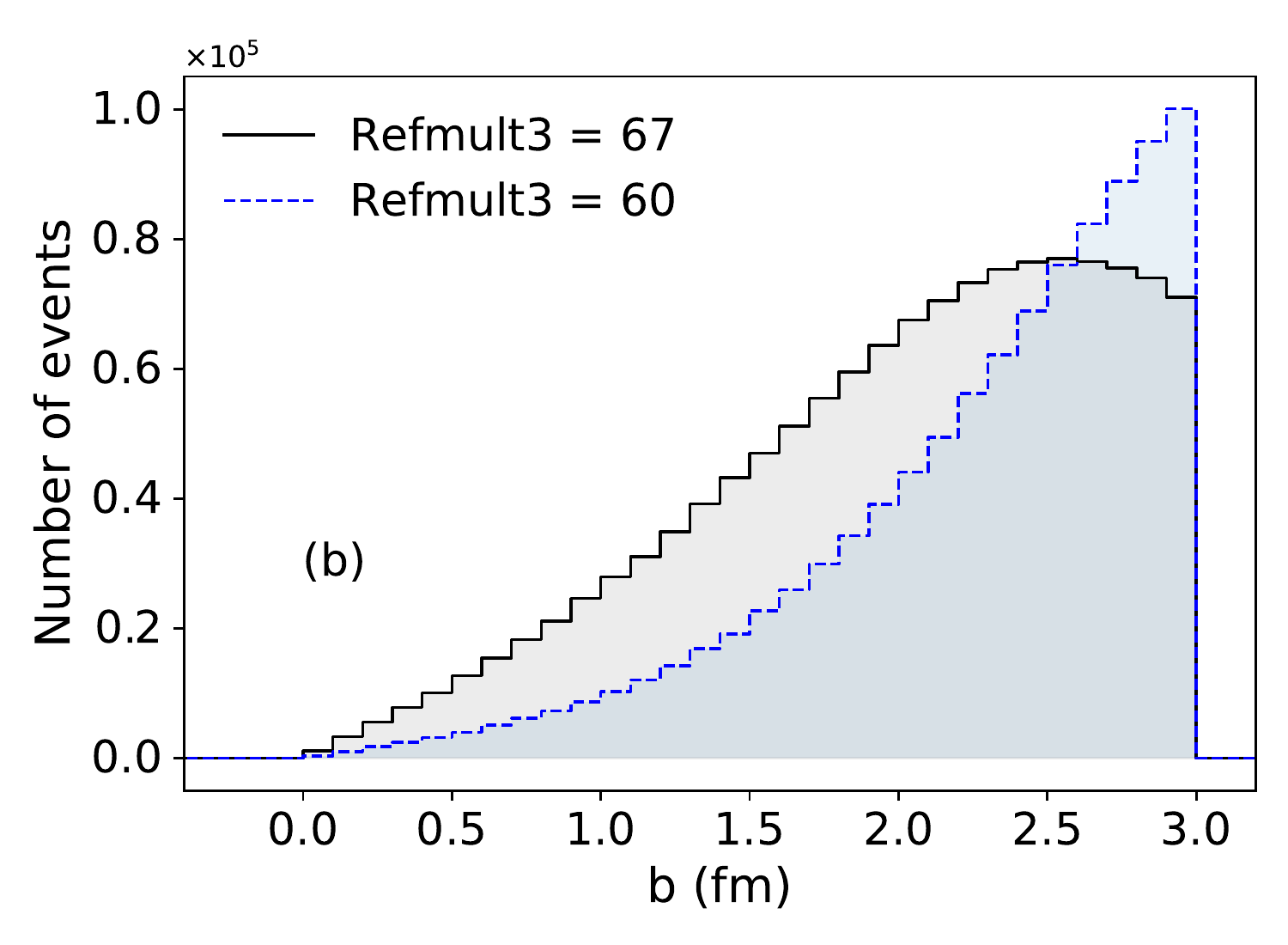}}
	\caption{(a) (Color online) Correlation between Refmult3 (charged particles excluding protons within $|\eta|<$0.5) and impact parameter in top central ($b<3$ fm) Au+Au collisions at \sNN = 3 GeV. (b) (Color online) Impact parameter distributions for fixed Refmult3 values.}
	\label{correlation}
\end{figure}
\section{Results}

In this section, we will start the discussion with proton $dN/dy$ distribution. Figure~\ref{rapidity} shows the $dN/dy$ distribution of proton and deuteron in most central Au+Au collisions at \sNN = 3 GeV in JAM model. The central collision are chosen with impact parameter less than 3 fm. The transverse momentum is set to be within $0.4<p_{T}<2.0$ GeV/$c$ for proton selection. The deuteron formation probability is proportional to the initial proton yield in that event according to coalescence after the kinetic freeze-out~\cite{Butler:1963pp, nagle1996coalescence}, i.e.  
\begin{eqnarray}
\lambda_{d} = Bn_{p_{i}}^{2},
\label{deuteron}
\end{eqnarray}
here we assume that the neutron yield is proportional to the proton yield in each event. $B$ and $n_{p_{i}}$ represents the coalescence parameter and initial proton number respectively. The above assumption is valid where volume fluctuation is minimum~\cite{Feckova:2015qza}. 
So the initial proton number (p without d-formation) can be approximated by adding observed proton and deuteron number as shown in Fig.~\ref{rapidity},
\begin{eqnarray}
\frac{dN_{p_{i}}}{dy} = n_{p_{i}} = n_{p} + n_{d}
\end{eqnarray}

Figure~\ref{netP} shows the event-by-event proton number distribution in most central Au+Au collision at \sNN = 3 GeV with and without deuteron formation in JAM model. The distributions are obtained by counting protons within $0.4<p_{T}<2.0$ GeV/$c$. The distributions presented in Fig.~\ref{netP} are not corrected by centrality bin width as described in previous section. 

Let us first discuss the validity of centrality bin width correction using Refmult3 at \sNN = 3 GeV. As we discussed in the previous section, at very low energies, even a single multiplicity bin corresponds to a wide initial volume fluctuation. This can be demonstrated in Fig.~\ref{correlation}. Figure~\ref{correlation}-(a) shows the two-dimension correlation plot between Refmult3 and impact parameter at \sNN = 3 GeV. We can observe that at 3 GeV, no such strong negative correlation is found between charged particles at mid-rapidity region (Refmult3) and impact parameter as observed in higher energies~\cite{Chatterjee:2019fey}. It indicates at low energies the charged particles at mid-rapidity region are insensitive to the initial collision geometry and have poor centrality resolution. Figure~\ref{correlation}-(b) shows the b-distributions for two different fixed Refmult3 values, 67 (peak value of Refmult3 distribution, have maximum weight) and 60. We can clearly see that even a fixed Refmult3 corresponds to all the impact parameter values from 0-3 fm with an almost similar weight as unbiased b-distribution.

\begin{figure*}[htp!]
	\centering 
	\includegraphics[width=0.8\textwidth]{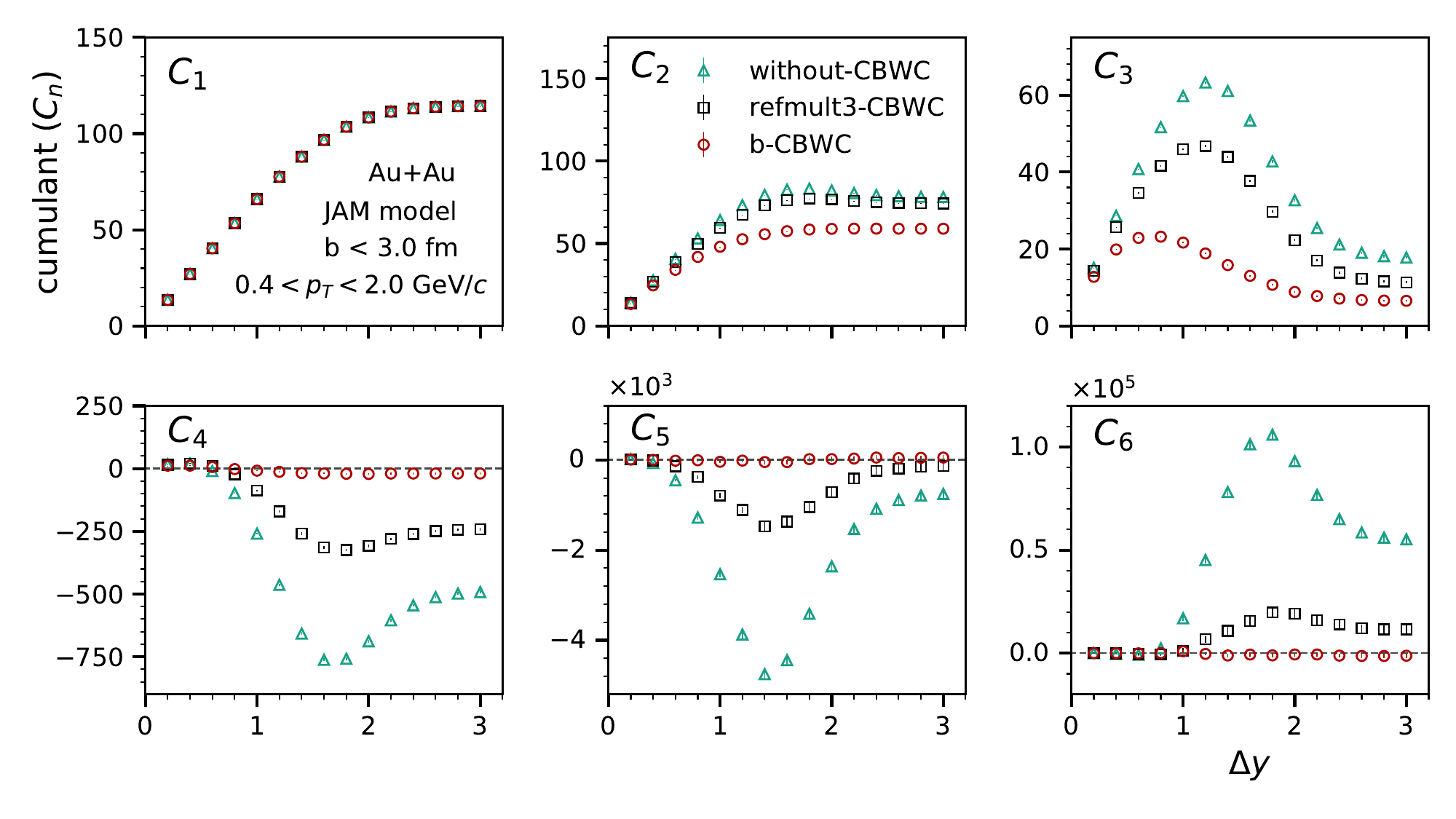}
	\caption{(Color online) Rapidity acceptance dependence cumulants of proton multiplicity distributions in top central ($b < 3$ fm) Au+Au collisions at \sNN = 3 GeV. The centrality bin width correction is done with (a) each Refmult3-bin (black square) and (b) 0.1 fm impact parameter bin (red circle). The results also compared with the cumulants calculated without CBWC (green triangle).}
	\label{cumulantcomparison}
\end{figure*}
\begin{figure*}[htp!]
	\centering 
	\includegraphics[width=0.8\textwidth]{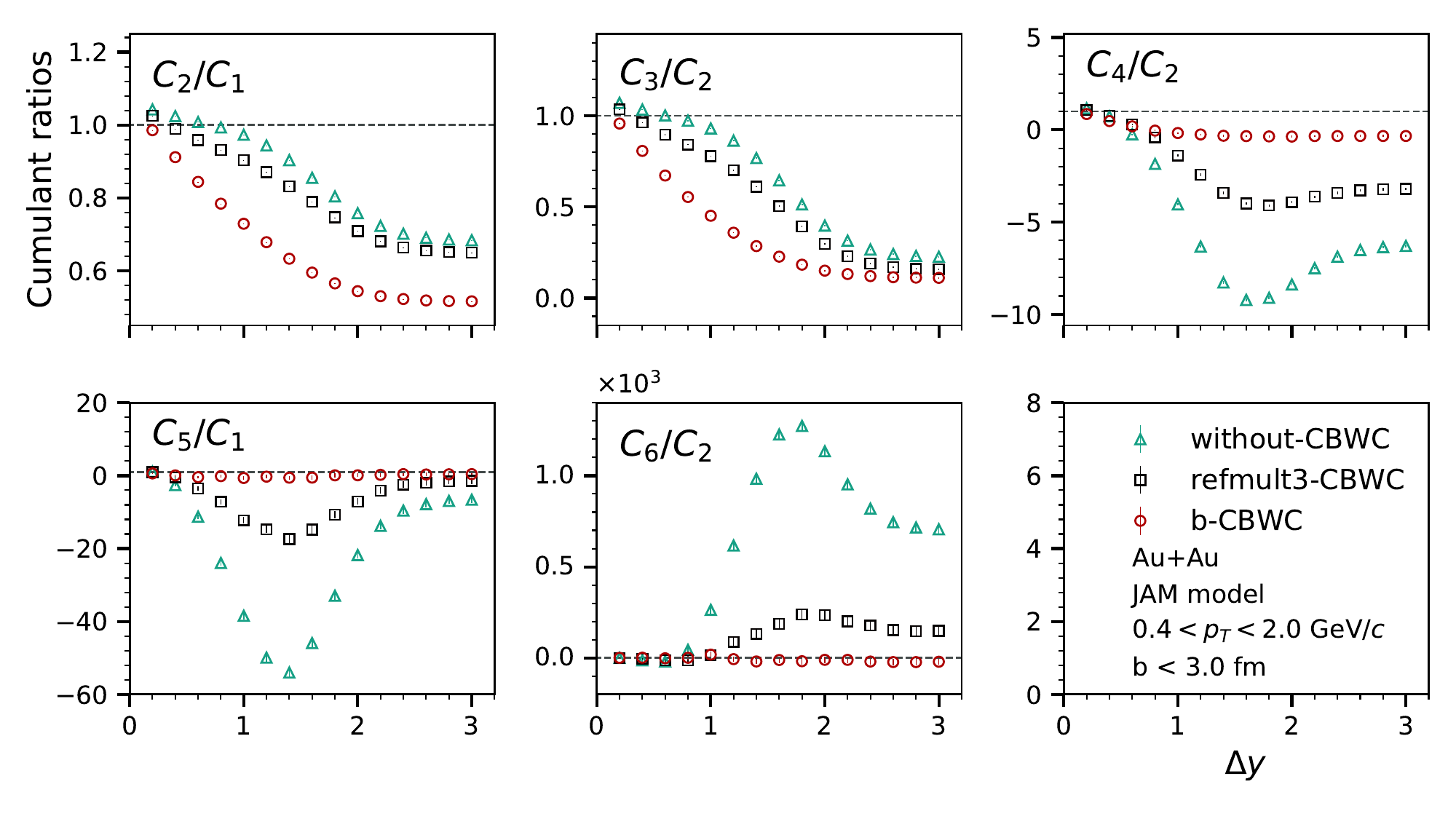}
	\caption{(Color online) Rapidity acceptance dependence cumulant ratios of proton multiplicity distributions in top central Au+Au collisions at \sNN = 3 GeV. The centrality bin width correction is done with (a) each Refmult3-bin (black square) and (b) 0.1 fm impact parameter bin (red circle). The results also compared with the cumulants calculated without CBWC (green triangle).}
	\label{ratiocomparison}
\end{figure*}

Figures~\ref{cumulantcomparison} and~\ref{ratiocomparison} show the rapidity acceptance dependence for the cumulants and cumulant ratios of proton multiplicity distributions for two different centrality definition in Au+Au collisions at \sNN = 3 GeV from the JAM model simulation. The results are also compared with the cumulants calculated without CBWC. We observed that both the cumulants and cumulant ratios obtained from Refmult3-CBWC have a large deviation from impact parameter CBWC at \sNN = 3 GeV. We used 0.1 fm bin for the impact parameter based CBWC.  From the above comparison we can conclude that unlike higher collision energies, the CBWC using charged-particle multiplicity bin cannot effectively suppress initial volume fluctuations in Au+Au collisions at \sNN = 3 GeV~\cite{Chatterjee:2019fey}. Thus, new methods for classifying events at low energy heavy-ion collisions are needed to determine the collision centralities.  Recently, in Refs.~\cite{Li:2020qqn,Kuttan:2020kha}, machine learning have been proposed to determine the collision centrality with high resolution in heavy-ion collisions. Those could be used to address the centrality fluctuation effect on cumulant analysis at low energies. In the subsequent sections, we use b-CBWC to understand the effect of deuteron formation on proton number cumulant and correlation functions.  

\begin{figure*}[htp!]
	\centering 
	\includegraphics[width=0.8\textwidth]{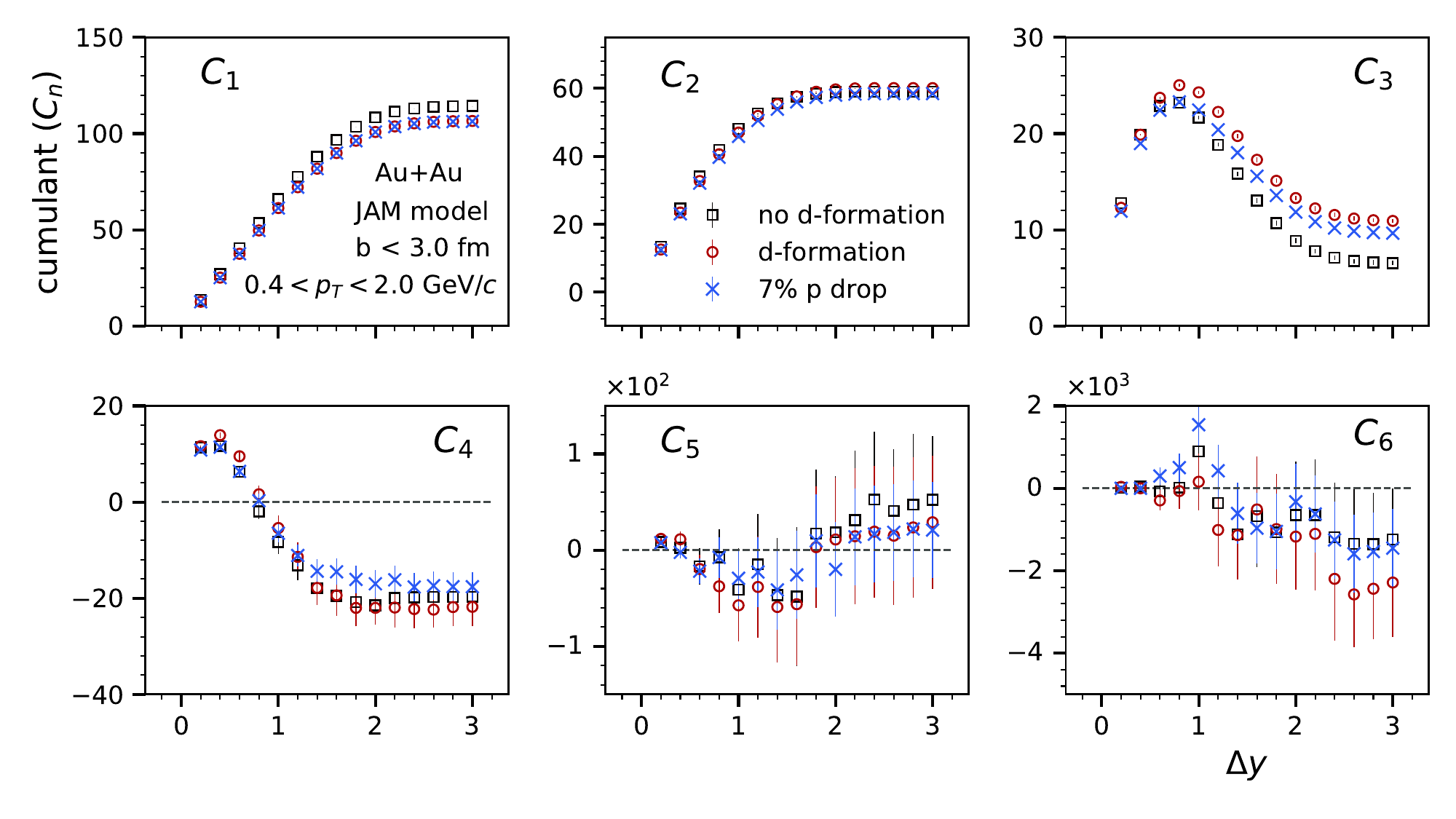}
	\caption{(Color online) Rapidity acceptance dependence cumulants ($C_{1} \sim C_{6}$) of proton multiplicity distributions in top central ($b < 3$ fm) Au+Au collisions at \sNN = 3 GeV. The results are obtained with/without deuteron formation in JAM model. The blue cross marker represents the $7\%$ random dropping of protons in each event}
	\label{cumulant}
\end{figure*}

Theoretically, it was predicted that the rapidity window dependence of proton cumulants are important observables to search for the QCD critical point and understand the non-equilibrium effects of dynamical expansion on the fluctuations in heavy-ion collisions~\cite{Ling:2015yau}. It is expected that the proton cumulant and correlation functions will shows power law dependence with the rapidity acceptance and number of protons as $C_{n},\kappa_{n} \propto (\Delta y)^{n} \propto (N_{p})^{n}$ due to the long range correlation close to the critical point. This relationship will holds if the rapidity acceptance is less than the typical correlation length near critical point ($\Delta y < \xi$)~\cite{Ling:2015yau, Zhang:2019lqz}. On the other hand, if the rapidity acceptance is large enough comparing to correlation length ($\Delta y \gg \xi$), the proton cumulant and multi-particle correlation function will be dominated by statistical fluctuation as $C_{n},\kappa_{n} \propto \Delta y \propto N_{p}$. However, if the rapidity acceptance is further enlarged the baryon number conservation effect will dominate over statistical fluctuation. 

\begin{figure*}[htp!]
	\centering 
	\includegraphics[width=0.8\textwidth]{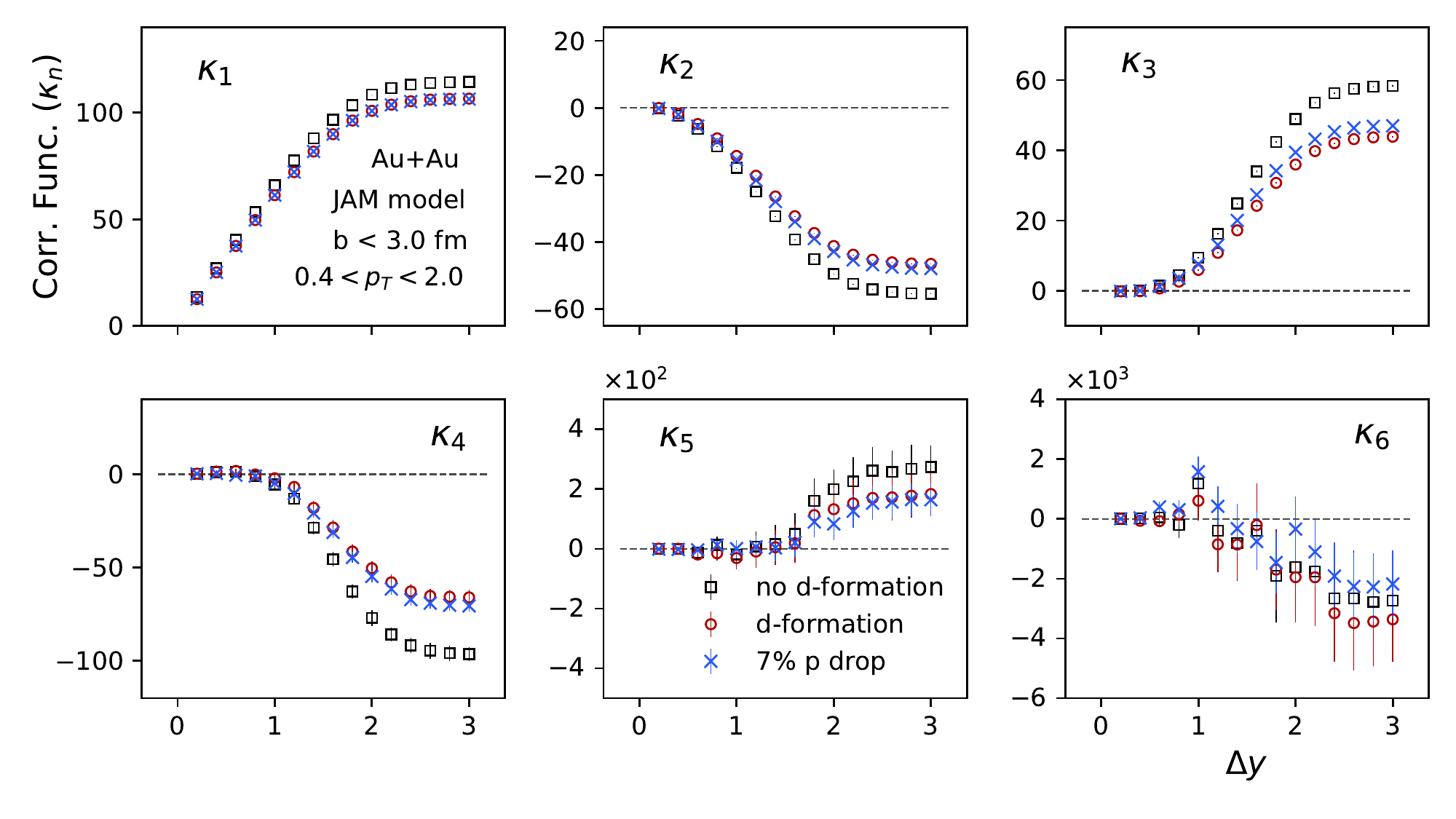}
	\caption{(Color online) Rapidity acceptance dependence of correlation functions ($\kappa_{1} \sim \kappa_{6}$) of proton multiplicity distribution in top central ($b < 3$ fm) Au+Au collision at \sNN = 3 GeV with/without deuteron formation. The blue cross marker represents the $7\%$ random dropping of protons in each event.}
	\label{correlationfunction}
\end{figure*}
\begin{figure*}[htp!]
	\centering 
	\includegraphics[width=0.8\textwidth]{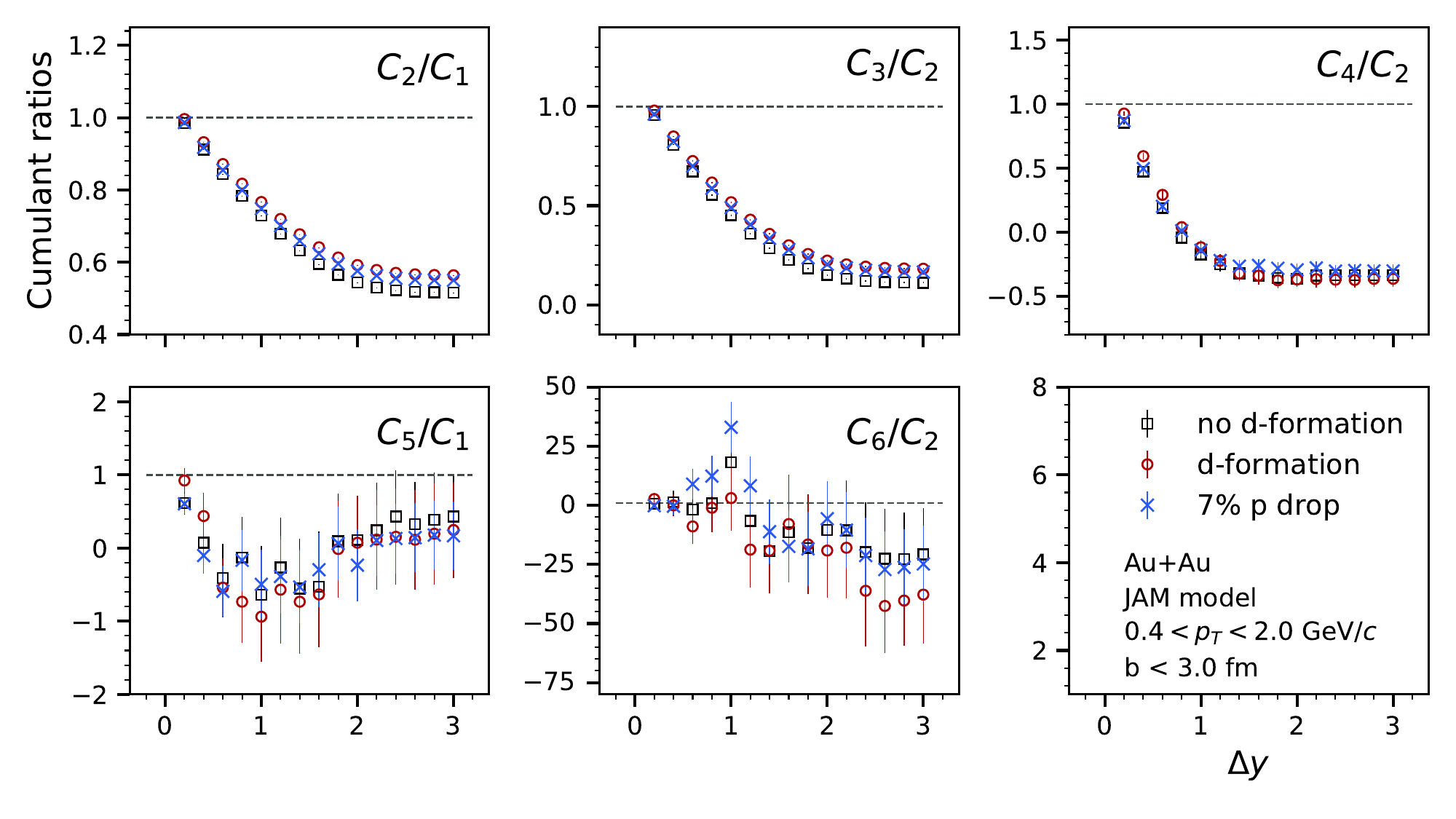}
	\caption{(Color online) Rapidity acceptance dependence cumulant ratios ($C_{2}/C_{1}$, $C_{3}/C_{2}$, $C_{4}/C_{2}$, $C_{5}/C_{1}$ and $C_{6}/C_{2}$) of proton multiplicity distributions in most central ($b < 3$ fm) Au+Au collisions at \sNN = 3 GeV. The results are obtained with/without deuteron formation in JAM model. The blue cross marker represents the $7\%$ random dropping of protons in each event.}
	\label{curatio}
\end{figure*}

Figure~\ref{cumulant} shows the variation of cumulants $C_{n}$ with the rapidity acceptance ($-y_{max} < y < y_{max}, ~\Delta y$ = 2 $y_{max}$) of proton multiplicity distributions in most central Au+Au collisions at \sNN = 3 GeV. The measurements are done within the transverse momentum range 0.4 to 2.0 GeV/$c$. All cumulants are saturate around $\Delta y \sim 2.2$, which is the acceptance up to beam rapidity ($y_{beam} = 1.039$ at \sNN = 3GeV)~\cite{Chatterjee:2019fey}. $C_{1}$ and $C_{2}$ linearly increase as a function of rapidity acceptance up to $2y_{beam}$ due to increase in proton number with acceptance. 
We observed about $7\%$ reduction in the mean value of protons for the case of deuteron formation. 
$C_{3}$ increases in low rapidity acceptance and shows a peak around $\Delta y \sim$ 0.7-0.8 and then decrease. Fourth order proton cumulant ($C_4$) values are negative above $\Delta y \sim$ 0.6 whereas fifth and sixth order proton cumulants  ($C_5$ and $C_6$) are consistent with zero with large statistical uncertainties. 
To obtain further understanding on effect of deuteron formation, we randomly reduced $7\%$ of total proton number in each event using binomial sampling. Although the values are not identical, we observed the effect of 7\% random dropping of protons are very close to the case of deuteron formation. 

Figure~\ref{correlationfunction} shows the rapidity acceptance dependence of correlation function $\kappa_{n}$ of protons in most central Au+Au collisions at \sNN = 3 GeV within the $p_{T}$ range 0.4 to 2.0 GeV/$c$. Different orders of correlation function values are saturate around $\Delta y \sim$ 2.2 around mid-rapidity. $\kappa_{1}$ increases as a function of rapidity window. The two particle correlation function ($\kappa_{2}$) of protons are found to be negative and decreases monotonically up to $\Delta y \sim 2.2$. 
Three particle correlation function  ($\kappa_{3}$) of proton increases with $\Delta y \sim$ acceptance. Fourth, fifth and sixth order correlation function of protons ($\kappa_{4}$,  $\kappa_{5}$ and $\kappa_{6}$)  are found to be close to zero up to $\Delta y \sim 1$, and start to deviate from zero when further enlarging the rapidity acceptance. Interestingly, the odd order correlation functions are found to be positive while even order correlation functions show negative values up to sixth order at large rapidity acceptance at \sNN = 3 GeV. Those strong rapidity acceptance dependence are mainly attributed to the effects of baryon number conservations~\cite{He:2017zpg,Poberezhnyuk:2020ayn,Braun-Munzinger:2020jbk,Pratt:2020ekp}. In addition, we observed that if we randomly drop $7\%$ protons in each event then the cumulants are close to the deuteron formation case.

Figure~\ref{curatio} shows rapidity acceptance dependence of cumulant ratios $C_{2}/C_{1}$,  $C_{3}/C_{2}$, $C_{4}/C_{2}$, $C_{5}/C_{1}$ and $C_{6}/C_{2}$ of proton multiplicity distributions in Au+Au collision at \sNN = 3 GeV. At small rapidity acceptance, the cumulant ratios follow statistical fluctuations (Poisson) baseline and the values are close to unity ($C_{m}/C_{n} \sim 1$). $C_{2}/C_{1}$ and $C_{3}/C_{2}$ decrease smoothly with $\Delta y$ and saturate around $\Delta y \sim 2$. The values of $C_{4}/C_{2}$ and $C_{5}/C_{1}$ are positive in small rapidity acceptance, then change sign around $\Delta y \sim 0.6$ and further decreases up to $\Delta y \sim 1$. $C_{6}/C_{2}$ is close to 1 within statistical uncertainty at smaller rapidity acceptance and shows negative values at larger acceptance. We observed a good agreement in cumulant ratios for deuteron formation and random dropping of proton case. 

\section{Summary}
In this work, we studied the effects of centrality fluctuation and deuteron formation on the cumulant and correlation function of protons up to sixth order in most central Au+Au collisions at \sNN = 3 GeV using JAM model. We presented the results as a function of the rapidity acceptance within transverse momentum $0.4<p_{T}<2 $ GeV/$c$. The proton cumulants  $C_{1}$ and $C_{2}$ increase linearly as a function of rapidity acceptance up to $\Delta y \sim 2y_{beam}$. It also leads to the suppression of the $C_3$ and $C_{4}$ at larger rapidity acceptance window ($\Delta y > 0.6$). Further, we found that the odd order correlation functions are found positive while even order correlation functions show negative values up to sixth order for larger rapidity acceptance at \sNN = 3 GeV. This is mainly due to the effects of baryon number conservation in heavy-ion collisions. The results obtained by centrality bin width correction (CBWC) using charged reference particle multiplicity are compared with the CBWC done using finer impact parameter bins. It was observed that the centrality resolution for determining the collision centrality using charged particle multiplicities cannot effectively reduce the centrality fluctuations in heavy-ion collisions at low energies. It brings challenges for us to carry out cumulant measurements in low energy heavy-ion collisions. New methods, such as machine learning techniques,  need to be built up  and applied to determine the collision centrality with high resolution, which is crucial for precisely measuring the higher-order cumulant  in heavy-ion collisions at low energies. 
On the other hand, we discussed the effect of deuteron formation on cumulant and correlation functions of protons and found it is similar to the binomial efficiency effect due to the loss of protons via deuteron formation. 
This work can serve as a non-critical  baselines  for  the  future  QCD  critical  point search in heavy-ion collisions at high baryon density region. 

\section{Acknowledgement}
X. Luo is grateful for the stimulating discussion with Dr. Jiangyong Jia and Dr. Nu Xu. This work is supported by the National Key Research and Development Program of China (Grant No. 2018YFE0205201), the National Natural Science Foundation of China (Grant No. 11828501, 11575069, 11890711 and 11861131009).

\bibliography{jamStudy}

\begin{thebibliography}{67}%
\makeatletter
\providecommand \@ifxundefined [1]{%
 \@ifx{#1\undefined}
}%
\providecommand \@ifnum [1]{%
 \ifnum #1\expandafter \@firstoftwo
 \else \expandafter \@secondoftwo
 \fi
}%
\providecommand \@ifx [1]{%
 \ifx #1\expandafter \@firstoftwo
 \else \expandafter \@secondoftwo
 \fi
}%
\providecommand \natexlab [1]{#1}%
\providecommand \enquote  [1]{``#1''}%
\providecommand \bibnamefont  [1]{#1}%
\providecommand \bibfnamefont [1]{#1}%
\providecommand \citenamefont [1]{#1}%
\providecommand \href@noop [0]{\@secondoftwo}%
\providecommand \href [0]{\begingroup \@sanitize@url \@href}%
\providecommand \@href[1]{\@@startlink{#1}\@@href}%
\providecommand \@@href[1]{\endgroup#1\@@endlink}%
\providecommand \@sanitize@url [0]{\catcode `\\12\catcode `\$12\catcode
  `\&12\catcode `\#12\catcode `\^12\catcode `\_12\catcode `\%12\relax}%
\providecommand \@@startlink[1]{}%
\providecommand \@@endlink[0]{}%
\providecommand \url  [0]{\begingroup\@sanitize@url \@url }%
\providecommand \@url [1]{\endgroup\@href {#1}{\urlprefix }}%
\providecommand \urlprefix  [0]{URL }%
\providecommand \Eprint [0]{\href }%
\providecommand \doibase [0]{http://dx.doi.org/}%
\providecommand \selectlanguage [0]{\@gobble}%
\providecommand \bibinfo  [0]{\@secondoftwo}%
\providecommand \bibfield  [0]{\@secondoftwo}%
\providecommand \translation [1]{[#1]}%
\providecommand \BibitemOpen [0]{}%
\providecommand \bibitemStop [0]{}%
\providecommand \bibitemNoStop [0]{.\EOS\space}%
\providecommand \EOS [0]{\spacefactor3000\relax}%
\providecommand \BibitemShut  [1]{\csname bibitem#1\endcsname}%
\let\auto@bib@innerbib\@empty
\bibitem [{\citenamefont {Aggarwal}\ \emph
  {et~al.}(2010{\natexlab{a}})\citenamefont {Aggarwal} \emph
  {et~al.}}]{Aggarwal:2010cw}%
  \BibitemOpen
  \bibfield  {author} {\bibinfo {author} {\bibfnamefont {M.~M.}\ \bibnamefont
  {Aggarwal}} \emph {et~al.} (\bibinfo {collaboration} {STAR}),\ }\href@noop {}
  {\  (\bibinfo {year} {2010}{\natexlab{a}})},\ \Eprint
  {http://arxiv.org/abs/1007.2613} {arXiv:1007.2613 [nucl-ex]} \BibitemShut
  {NoStop}%
\bibitem [{\citenamefont {Svetitsky}(1986)}]{Svetitsky:1985ye}%
  \BibitemOpen
  \bibfield  {author} {\bibinfo {author} {\bibfnamefont {B.}~\bibnamefont
  {Svetitsky}},\ }\href {\doibase 10.1016/0370-1573(86)90014-1} {\bibfield
  {journal} {\bibinfo  {journal} {Phys. Rept.}\ }\textbf {\bibinfo {volume}
  {132}},\ \bibinfo {pages} {1} (\bibinfo {year} {1986})}\BibitemShut {NoStop}%
\bibitem [{\citenamefont {Aoki}\ \emph {et~al.}(2006)\citenamefont {Aoki},
  \citenamefont {Endrodi}, \citenamefont {Fodor}, \citenamefont {Katz},\ and\
  \citenamefont {Szabo}}]{Aoki:2006we}%
  \BibitemOpen
  \bibfield  {author} {\bibinfo {author} {\bibfnamefont {Y.}~\bibnamefont
  {Aoki}}, \bibinfo {author} {\bibfnamefont {G.}~\bibnamefont {Endrodi}},
  \bibinfo {author} {\bibfnamefont {Z.}~\bibnamefont {Fodor}}, \bibinfo
  {author} {\bibfnamefont {S.~D.}\ \bibnamefont {Katz}}, \ and\ \bibinfo
  {author} {\bibfnamefont {K.~K.}\ \bibnamefont {Szabo}},\ }\href {\doibase
  10.1038/nature05120} {\bibfield  {journal} {\bibinfo  {journal} {Nature}\
  }\textbf {\bibinfo {volume} {443}},\ \bibinfo {pages} {675} (\bibinfo {year}
  {2006})},\ \Eprint {http://arxiv.org/abs/hep-lat/0611014}
  {arXiv:hep-lat/0611014 [hep-lat]} \BibitemShut {NoStop}%
\bibitem [{\citenamefont {Gupta}\ \emph {et~al.}(2011)\citenamefont {Gupta},
  \citenamefont {Luo}, \citenamefont {Mohanty}, \citenamefont {Ritter},\ and\
  \citenamefont {Xu}}]{Gupta:2011wh}%
  \BibitemOpen
  \bibfield  {author} {\bibinfo {author} {\bibfnamefont {S.}~\bibnamefont
  {Gupta}}, \bibinfo {author} {\bibfnamefont {X.}~\bibnamefont {Luo}}, \bibinfo
  {author} {\bibfnamefont {B.}~\bibnamefont {Mohanty}}, \bibinfo {author}
  {\bibfnamefont {H.~G.}\ \bibnamefont {Ritter}}, \ and\ \bibinfo {author}
  {\bibfnamefont {N.}~\bibnamefont {Xu}},\ }\href {\doibase
  10.1126/science.1204621} {\bibfield  {journal} {\bibinfo  {journal}
  {Science}\ }\textbf {\bibinfo {volume} {332}},\ \bibinfo {pages} {1525}
  (\bibinfo {year} {2011})},\ \Eprint {http://arxiv.org/abs/1105.3934}
  {arXiv:1105.3934 [hep-ph]} \BibitemShut {NoStop}%
\bibitem [{\citenamefont {Fodor}\ and\ \citenamefont
  {Katz}(2004)}]{Fodor:2004nz}%
  \BibitemOpen
  \bibfield  {author} {\bibinfo {author} {\bibfnamefont {Z.}~\bibnamefont
  {Fodor}}\ and\ \bibinfo {author} {\bibfnamefont {S.~D.}\ \bibnamefont
  {Katz}},\ }\href {\doibase 10.1088/1126-6708/2004/04/050} {\bibfield
  {journal} {\bibinfo  {journal} {JHEP}\ }\textbf {\bibinfo {volume} {04}},\
  \bibinfo {pages} {050} (\bibinfo {year} {2004})},\ \Eprint
  {http://arxiv.org/abs/hep-lat/0402006} {arXiv:hep-lat/0402006 [hep-lat]}
  \BibitemShut {NoStop}%
\bibitem [{\citenamefont {de~Forcrand}\ and\ \citenamefont
  {Philipsen}(2002)}]{deForcrand:2002hgr}%
  \BibitemOpen
  \bibfield  {author} {\bibinfo {author} {\bibfnamefont {P.}~\bibnamefont
  {de~Forcrand}}\ and\ \bibinfo {author} {\bibfnamefont {O.}~\bibnamefont
  {Philipsen}},\ }\href {\doibase 10.1016/S0550-3213(02)00626-0} {\bibfield
  {journal} {\bibinfo  {journal} {Nucl. Phys.}\ }\textbf {\bibinfo {volume}
  {B642}},\ \bibinfo {pages} {290} (\bibinfo {year} {2002})},\ \Eprint
  {http://arxiv.org/abs/hep-lat/0205016} {arXiv:hep-lat/0205016 [hep-lat]}
  \BibitemShut {NoStop}%
\bibitem [{\citenamefont {Qin}\ \emph {et~al.}(2011)\citenamefont {Qin},
  \citenamefont {Chang}, \citenamefont {Chen}, \citenamefont {Liu},\ and\
  \citenamefont {Roberts}}]{Qin:2010nq}%
  \BibitemOpen
  \bibfield  {author} {\bibinfo {author} {\bibfnamefont {S.-x.}\ \bibnamefont
  {Qin}}, \bibinfo {author} {\bibfnamefont {L.}~\bibnamefont {Chang}}, \bibinfo
  {author} {\bibfnamefont {H.}~\bibnamefont {Chen}}, \bibinfo {author}
  {\bibfnamefont {Y.-x.}\ \bibnamefont {Liu}}, \ and\ \bibinfo {author}
  {\bibfnamefont {C.~D.}\ \bibnamefont {Roberts}},\ }\href {\doibase
  10.1103/PhysRevLett.106.172301} {\bibfield  {journal} {\bibinfo  {journal}
  {Phys. Rev. Lett.}\ }\textbf {\bibinfo {volume} {106}},\ \bibinfo {pages}
  {172301} (\bibinfo {year} {2011})},\ \Eprint {http://arxiv.org/abs/1011.2876}
  {arXiv:1011.2876 [nucl-th]} \BibitemShut {NoStop}%
\bibitem [{\citenamefont {Xin}\ \emph {et~al.}(2014)\citenamefont {Xin},
  \citenamefont {Qin},\ and\ \citenamefont {Liu}}]{Xin:2014ela}%
  \BibitemOpen
  \bibfield  {author} {\bibinfo {author} {\bibfnamefont {X.-y.}\ \bibnamefont
  {Xin}}, \bibinfo {author} {\bibfnamefont {S.-x.}\ \bibnamefont {Qin}}, \ and\
  \bibinfo {author} {\bibfnamefont {Y.-x.}\ \bibnamefont {Liu}},\ }\href
  {\doibase 10.1103/PhysRevD.90.076006} {\bibfield  {journal} {\bibinfo
  {journal} {Phys. Rev.}\ }\textbf {\bibinfo {volume} {D90}},\ \bibinfo {pages}
  {076006} (\bibinfo {year} {2014})}\BibitemShut {NoStop}%
\bibitem [{\citenamefont {Shi}\ \emph {et~al.}(2014)\citenamefont {Shi},
  \citenamefont {Wang}, \citenamefont {Jiang}, \citenamefont {Cui},\ and\
  \citenamefont {Zong}}]{Shi:2014zpa}%
  \BibitemOpen
  \bibfield  {author} {\bibinfo {author} {\bibfnamefont {C.}~\bibnamefont
  {Shi}}, \bibinfo {author} {\bibfnamefont {Y.-L.}\ \bibnamefont {Wang}},
  \bibinfo {author} {\bibfnamefont {Y.}~\bibnamefont {Jiang}}, \bibinfo
  {author} {\bibfnamefont {Z.-F.}\ \bibnamefont {Cui}}, \ and\ \bibinfo
  {author} {\bibfnamefont {H.-S.}\ \bibnamefont {Zong}},\ }\href {\doibase
  10.1007/JHEP07(2014)014} {\bibfield  {journal} {\bibinfo  {journal} {JHEP}\
  }\textbf {\bibinfo {volume} {07}},\ \bibinfo {pages} {014} (\bibinfo {year}
  {2014})},\ \Eprint {http://arxiv.org/abs/1403.3797} {arXiv:1403.3797
  [hep-ph]} \BibitemShut {NoStop}%
\bibitem [{\citenamefont {Fischer}\ \emph {et~al.}(2014)\citenamefont
  {Fischer}, \citenamefont {Luecker},\ and\ \citenamefont
  {Welzbacher}}]{Fischer:2014ata}%
  \BibitemOpen
  \bibfield  {author} {\bibinfo {author} {\bibfnamefont {C.~S.}\ \bibnamefont
  {Fischer}}, \bibinfo {author} {\bibfnamefont {J.}~\bibnamefont {Luecker}}, \
  and\ \bibinfo {author} {\bibfnamefont {C.~A.}\ \bibnamefont {Welzbacher}},\
  }\href {\doibase 10.1103/PhysRevD.90.034022} {\bibfield  {journal} {\bibinfo
  {journal} {Phys. Rev.}\ }\textbf {\bibinfo {volume} {D90}},\ \bibinfo {pages}
  {034022} (\bibinfo {year} {2014})},\ \Eprint {http://arxiv.org/abs/1405.4762}
  {arXiv:1405.4762 [hep-ph]} \BibitemShut {NoStop}%
\bibitem [{\citenamefont {Lu}\ \emph {et~al.}(2015)\citenamefont {Lu},
  \citenamefont {Du}, \citenamefont {Cui},\ and\ \citenamefont
  {Zong}}]{Lu:2015naa}%
  \BibitemOpen
  \bibfield  {author} {\bibinfo {author} {\bibfnamefont {Y.}~\bibnamefont
  {Lu}}, \bibinfo {author} {\bibfnamefont {Y.-L.}\ \bibnamefont {Du}}, \bibinfo
  {author} {\bibfnamefont {Z.-F.}\ \bibnamefont {Cui}}, \ and\ \bibinfo
  {author} {\bibfnamefont {H.-S.}\ \bibnamefont {Zong}},\ }\href {\doibase
  10.1140/epjc/s10052-015-3720-2} {\bibfield  {journal} {\bibinfo  {journal}
  {Eur. Phys. J.}\ }\textbf {\bibinfo {volume} {C75}},\ \bibinfo {pages} {495}
  (\bibinfo {year} {2015})},\ \Eprint {http://arxiv.org/abs/1508.00651}
  {arXiv:1508.00651 [hep-ph]} \BibitemShut {NoStop}%
\bibitem [{\citenamefont {Bazavov}\ \emph {et~al.}(2017)\citenamefont {Bazavov}
  \emph {et~al.}}]{Bazavov:2017tot}%
  \BibitemOpen
  \bibfield  {author} {\bibinfo {author} {\bibfnamefont {A.}~\bibnamefont
  {Bazavov}} \emph {et~al.} (\bibinfo {collaboration} {HotQCD}),\ }\href
  {\doibase 10.1103/PhysRevD.96.074510} {\bibfield  {journal} {\bibinfo
  {journal} {Phys. Rev.}\ }\textbf {\bibinfo {volume} {D96}},\ \bibinfo {pages}
  {074510} (\bibinfo {year} {2017})},\ \Eprint
  {http://arxiv.org/abs/1708.04897} {arXiv:1708.04897 [hep-lat]} \BibitemShut
  {NoStop}%
\bibitem [{\citenamefont {Fu}\ \emph {et~al.}(2019)\citenamefont {Fu},
  \citenamefont {Pawlowski},\ and\ \citenamefont {Rennecke}}]{Fu:2019hdw}%
  \BibitemOpen
  \bibfield  {author} {\bibinfo {author} {\bibfnamefont {W.-j.}\ \bibnamefont
  {Fu}}, \bibinfo {author} {\bibfnamefont {J.~M.}\ \bibnamefont {Pawlowski}}, \
  and\ \bibinfo {author} {\bibfnamefont {F.}~\bibnamefont {Rennecke}},\
  }\href@noop {} {\  (\bibinfo {year} {2019})},\ \Eprint
  {http://arxiv.org/abs/1909.02991} {arXiv:1909.02991 [hep-ph]} \BibitemShut
  {NoStop}%
\bibitem [{\citenamefont {Fischer}(2019)}]{Fischer:2018sdj}%
  \BibitemOpen
  \bibfield  {author} {\bibinfo {author} {\bibfnamefont {C.~S.}\ \bibnamefont
  {Fischer}},\ }\href {\doibase 10.1016/j.ppnp.2019.01.002} {\bibfield
  {journal} {\bibinfo  {journal} {Prog. Part. Nucl. Phys.}\ }\textbf {\bibinfo
  {volume} {105}},\ \bibinfo {pages} {1} (\bibinfo {year} {2019})},\ \Eprint
  {http://arxiv.org/abs/1810.12938} {arXiv:1810.12938 [hep-ph]} \BibitemShut
  {NoStop}%
\bibitem [{\citenamefont {Li}\ \emph {et~al.}(2019)\citenamefont {Li},
  \citenamefont {Xu}, \citenamefont {Wang},\ and\ \citenamefont
  {Huang}}]{Li:2018ygx}%
  \BibitemOpen
  \bibfield  {author} {\bibinfo {author} {\bibfnamefont {Z.}~\bibnamefont
  {Li}}, \bibinfo {author} {\bibfnamefont {K.}~\bibnamefont {Xu}}, \bibinfo
  {author} {\bibfnamefont {X.}~\bibnamefont {Wang}}, \ and\ \bibinfo {author}
  {\bibfnamefont {M.}~\bibnamefont {Huang}},\ }\href {\doibase
  10.1140/epjc/s10052-019-6703-x} {\bibfield  {journal} {\bibinfo  {journal}
  {Eur. Phys. J.}\ }\textbf {\bibinfo {volume} {C79}},\ \bibinfo {pages} {245}
  (\bibinfo {year} {2019})},\ \Eprint {http://arxiv.org/abs/1801.09215}
  {arXiv:1801.09215 [hep-ph]} \BibitemShut {NoStop}%
\bibitem [{\citenamefont {Yu}\ \emph {et~al.}(2020)\citenamefont {Yu},
  \citenamefont {Zhang},\ and\ \citenamefont {Luo}}]{Yu:2018kvh}%
  \BibitemOpen
  \bibfield  {author} {\bibinfo {author} {\bibfnamefont {N.}~\bibnamefont
  {Yu}}, \bibinfo {author} {\bibfnamefont {D.}~\bibnamefont {Zhang}}, \ and\
  \bibinfo {author} {\bibfnamefont {X.}~\bibnamefont {Luo}},\ }\href {\doibase
  10.1088/1674-1137/44/1/014002} {\bibfield  {journal} {\bibinfo  {journal}
  {Chin. Phys. C}\ }\textbf {\bibinfo {volume} {44}},\ \bibinfo {pages}
  {014002} (\bibinfo {year} {2020})},\ \Eprint
  {http://arxiv.org/abs/1812.04291} {arXiv:1812.04291 [nucl-th]} \BibitemShut
  {NoStop}%
\bibitem [{\citenamefont {Aggarwal}\ \emph
  {et~al.}(2010{\natexlab{b}})\citenamefont {Aggarwal} \emph
  {et~al.}}]{Aggarwal:2010wy}%
  \BibitemOpen
  \bibfield  {author} {\bibinfo {author} {\bibfnamefont {M.~M.}\ \bibnamefont
  {Aggarwal}} \emph {et~al.} (\bibinfo {collaboration} {STAR}),\ }\href
  {\doibase 10.1103/PhysRevLett.105.022302} {\bibfield  {journal} {\bibinfo
  {journal} {Phys. Rev. Lett.}\ }\textbf {\bibinfo {volume} {105}},\ \bibinfo
  {pages} {022302} (\bibinfo {year} {2010}{\natexlab{b}})},\ \Eprint
  {http://arxiv.org/abs/1004.4959} {arXiv:1004.4959 [nucl-ex]} \BibitemShut
  {NoStop}%
\bibitem [{\citenamefont {Adamczyk}\ \emph
  {et~al.}(2014{\natexlab{a}})\citenamefont {Adamczyk} \emph
  {et~al.}}]{Adamczyk:2013dal}%
  \BibitemOpen
  \bibfield  {author} {\bibinfo {author} {\bibfnamefont {L.}~\bibnamefont
  {Adamczyk}} \emph {et~al.} (\bibinfo {collaboration} {STAR}),\ }\href
  {\doibase 10.1103/PhysRevLett.112.032302} {\bibfield  {journal} {\bibinfo
  {journal} {Phys. Rev. Lett.}\ }\textbf {\bibinfo {volume} {112}},\ \bibinfo
  {pages} {032302} (\bibinfo {year} {2014}{\natexlab{a}})},\ \Eprint
  {http://arxiv.org/abs/1309.5681} {arXiv:1309.5681 [nucl-ex]} \BibitemShut
  {NoStop}%
\bibitem [{\citenamefont {Luo}(2016)}]{Luo:2015doi}%
  \BibitemOpen
  \bibfield  {author} {\bibinfo {author} {\bibfnamefont {X.}~\bibnamefont
  {Luo}},\ }\bibfield  {booktitle} {\emph {\bibinfo {booktitle} {{Proceedings,
  25th International Conference on Ultra-Relativistic Nucleus-Nucleus
  Collisions (Quark Matter 2015): Kobe, Japan, September 27-October 3,
  2015}}},\ }\href {\doibase 10.1016/j.nuclphysa.2016.03.025} {\bibfield
  {journal} {\bibinfo  {journal} {Nucl. Phys.}\ }\textbf {\bibinfo {volume}
  {A956}},\ \bibinfo {pages} {75} (\bibinfo {year} {2016})},\ \Eprint
  {http://arxiv.org/abs/1512.09215} {arXiv:1512.09215 [nucl-ex]} \BibitemShut
  {NoStop}%
\bibitem [{\citenamefont {Adamczyk}\ \emph
  {et~al.}(2014{\natexlab{b}})\citenamefont {Adamczyk} \emph
  {et~al.}}]{Adamczyk:2014fia}%
  \BibitemOpen
  \bibfield  {author} {\bibinfo {author} {\bibfnamefont {L.}~\bibnamefont
  {Adamczyk}} \emph {et~al.} (\bibinfo {collaboration} {STAR}),\ }\href
  {\doibase 10.1103/PhysRevLett.113.092301} {\bibfield  {journal} {\bibinfo
  {journal} {Phys. Rev. Lett.}\ }\textbf {\bibinfo {volume} {113}},\ \bibinfo
  {pages} {092301} (\bibinfo {year} {2014}{\natexlab{b}})},\ \Eprint
  {http://arxiv.org/abs/1402.1558} {arXiv:1402.1558 [nucl-ex]} \BibitemShut
  {NoStop}%
\bibitem [{\citenamefont {Adamczyk}\ \emph {et~al.}(2018)\citenamefont
  {Adamczyk} \emph {et~al.}}]{Adamczyk:2017wsl}%
  \BibitemOpen
  \bibfield  {author} {\bibinfo {author} {\bibfnamefont {L.}~\bibnamefont
  {Adamczyk}} \emph {et~al.} (\bibinfo {collaboration} {STAR}),\ }\href
  {\doibase 10.1016/j.physletb.2018.07.066} {\bibfield  {journal} {\bibinfo
  {journal} {Phys. Lett.}\ }\textbf {\bibinfo {volume} {B785}},\ \bibinfo
  {pages} {551} (\bibinfo {year} {2018})},\ \Eprint
  {http://arxiv.org/abs/1709.00773} {arXiv:1709.00773 [nucl-ex]} \BibitemShut
  {NoStop}%
\bibitem [{\citenamefont {Rajagopal}(1999)}]{Rajagopal:1999cp}%
  \BibitemOpen
  \bibfield  {author} {\bibinfo {author} {\bibfnamefont {K.}~\bibnamefont
  {Rajagopal}},\ }\bibfield  {booktitle} {\emph {\bibinfo {booktitle} {{Quark
  matter '99. Proceedings, 14th International Conference on ultrarelativistic
  nucleus nucleus collisions, QM'99, Torino, Italy, May 10-15, 1999}}},\ }\href
  {\doibase 10.1016/S0375-9474(99)85017-9} {\bibfield  {journal} {\bibinfo
  {journal} {Nucl. Phys.}\ }\textbf {\bibinfo {volume} {A661}},\ \bibinfo
  {pages} {150} (\bibinfo {year} {1999})},\ \Eprint
  {http://arxiv.org/abs/hep-ph/9908360} {arXiv:hep-ph/9908360 [hep-ph]}
  \BibitemShut {NoStop}%
\bibitem [{\citenamefont {Stephanov}\ \emph {et~al.}(1998)\citenamefont
  {Stephanov}, \citenamefont {Rajagopal},\ and\ \citenamefont
  {Shuryak}}]{Stephanov:1998dy}%
  \BibitemOpen
  \bibfield  {author} {\bibinfo {author} {\bibfnamefont {M.~A.}\ \bibnamefont
  {Stephanov}}, \bibinfo {author} {\bibfnamefont {K.}~\bibnamefont
  {Rajagopal}}, \ and\ \bibinfo {author} {\bibfnamefont {E.~V.}\ \bibnamefont
  {Shuryak}},\ }\href {\doibase 10.1103/PhysRevLett.81.4816} {\bibfield
  {journal} {\bibinfo  {journal} {Phys. Rev. Lett.}\ }\textbf {\bibinfo
  {volume} {81}},\ \bibinfo {pages} {4816} (\bibinfo {year} {1998})},\ \Eprint
  {http://arxiv.org/abs/hep-ph/9806219} {arXiv:hep-ph/9806219 [hep-ph]}
  \BibitemShut {NoStop}%
\bibitem [{\citenamefont {Luo}\ and\ \citenamefont {Xu}(2017)}]{Luo:2017faz}%
  \BibitemOpen
  \bibfield  {author} {\bibinfo {author} {\bibfnamefont {X.}~\bibnamefont
  {Luo}}\ and\ \bibinfo {author} {\bibfnamefont {N.}~\bibnamefont {Xu}},\
  }\href {\doibase 10.1007/s41365-017-0257-0} {\bibfield  {journal} {\bibinfo
  {journal} {Nucl. Sci. Tech.}\ }\textbf {\bibinfo {volume} {28}},\ \bibinfo
  {pages} {112} (\bibinfo {year} {2017})},\ \Eprint
  {http://arxiv.org/abs/1701.02105} {arXiv:1701.02105 [nucl-ex]} \BibitemShut
  {NoStop}%
\bibitem [{\citenamefont {Bzdak}\ \emph {et~al.}(2020)\citenamefont {Bzdak},
  \citenamefont {Esumi}, \citenamefont {Koch}, \citenamefont {Liao},
  \citenamefont {Stephanov},\ and\ \citenamefont {Xu}}]{Bzdak:2019pkr}%
  \BibitemOpen
  \bibfield  {author} {\bibinfo {author} {\bibfnamefont {A.}~\bibnamefont
  {Bzdak}}, \bibinfo {author} {\bibfnamefont {S.}~\bibnamefont {Esumi}},
  \bibinfo {author} {\bibfnamefont {V.}~\bibnamefont {Koch}}, \bibinfo {author}
  {\bibfnamefont {J.}~\bibnamefont {Liao}}, \bibinfo {author} {\bibfnamefont
  {M.}~\bibnamefont {Stephanov}}, \ and\ \bibinfo {author} {\bibfnamefont
  {N.}~\bibnamefont {Xu}},\ }\href {\doibase 10.1016/j.physrep.2020.01.005}
  {\bibfield  {journal} {\bibinfo  {journal} {Phys. Rept.}\ }\textbf {\bibinfo
  {volume} {853}},\ \bibinfo {pages} {1} (\bibinfo {year} {2020})},\ \Eprint
  {http://arxiv.org/abs/1906.00936} {arXiv:1906.00936 [nucl-th]} \BibitemShut
  {NoStop}%
\bibitem [{\citenamefont {Luo}\ \emph {et~al.}(2020)\citenamefont {Luo},
  \citenamefont {Shi}, \citenamefont {Xu},\ and\ \citenamefont
  {Zhang}}]{Luo:2020pef}%
  \BibitemOpen
  \bibfield  {author} {\bibinfo {author} {\bibfnamefont {X.}~\bibnamefont
  {Luo}}, \bibinfo {author} {\bibfnamefont {S.}~\bibnamefont {Shi}}, \bibinfo
  {author} {\bibfnamefont {N.}~\bibnamefont {Xu}}, \ and\ \bibinfo {author}
  {\bibfnamefont {Y.}~\bibnamefont {Zhang}},\ }\href {\doibase
  10.3390/particles3020022} {\bibfield  {journal} {\bibinfo  {journal}
  {Particles}\ }\textbf {\bibinfo {volume} {3}},\ \bibinfo {pages} {278}
  (\bibinfo {year} {2020})},\ \Eprint {http://arxiv.org/abs/2004.00789}
  {arXiv:2004.00789 [nucl-ex]} \BibitemShut {NoStop}%
\bibitem [{\citenamefont {Friman}\ \emph {et~al.}(2011)\citenamefont {Friman},
  \citenamefont {Karsch}, \citenamefont {Redlich},\ and\ \citenamefont
  {Skokov}}]{Friman:2011pf}%
  \BibitemOpen
  \bibfield  {author} {\bibinfo {author} {\bibfnamefont {B.}~\bibnamefont
  {Friman}}, \bibinfo {author} {\bibfnamefont {F.}~\bibnamefont {Karsch}},
  \bibinfo {author} {\bibfnamefont {K.}~\bibnamefont {Redlich}}, \ and\
  \bibinfo {author} {\bibfnamefont {V.}~\bibnamefont {Skokov}},\ }\href
  {\doibase 10.1140/epjc/s10052-011-1694-2} {\bibfield  {journal} {\bibinfo
  {journal} {Eur. Phys. J. C}\ }\textbf {\bibinfo {volume} {71}},\ \bibinfo
  {pages} {1694} (\bibinfo {year} {2011})},\ \Eprint
  {http://arxiv.org/abs/1103.3511} {arXiv:1103.3511 [hep-ph]} \BibitemShut
  {NoStop}%
\bibitem [{\citenamefont {Bazavov}\ \emph {et~al.}(2020)\citenamefont {Bazavov}
  \emph {et~al.}}]{Bazavov:2020bjn}%
  \BibitemOpen
  \bibfield  {author} {\bibinfo {author} {\bibfnamefont {A.}~\bibnamefont
  {Bazavov}} \emph {et~al.},\ }\href {\doibase 10.1103/PhysRevD.101.074502}
  {\bibfield  {journal} {\bibinfo  {journal} {Phys. Rev. D}\ }\textbf {\bibinfo
  {volume} {101}},\ \bibinfo {pages} {074502} (\bibinfo {year} {2020})},\
  \Eprint {http://arxiv.org/abs/2001.08530} {arXiv:2001.08530 [hep-lat]}
  \BibitemShut {NoStop}%
\bibitem [{\citenamefont {Nonaka}(2020)}]{Nonaka:2020crv}%
  \BibitemOpen
  \bibfield  {author} {\bibinfo {author} {\bibfnamefont {T.}~\bibnamefont
  {Nonaka}} (\bibinfo {collaboration} {STAR}),\ }in\ \href@noop {} {\emph
  {\bibinfo {booktitle} {{28th International Conference on Ultrarelativistic
  Nucleus-Nucleus Collisions}}}}\ (\bibinfo {year} {2020})\ \Eprint
  {http://arxiv.org/abs/2002.12505} {arXiv:2002.12505 [nucl-ex]} \BibitemShut
  {NoStop}%
\bibitem [{\citenamefont {Adare}\ \emph {et~al.}(2016)\citenamefont {Adare}
  \emph {et~al.}}]{Adare:2015aqk}%
  \BibitemOpen
  \bibfield  {author} {\bibinfo {author} {\bibfnamefont {A.}~\bibnamefont
  {Adare}} \emph {et~al.} (\bibinfo {collaboration} {PHENIX}),\ }\href
  {\doibase 10.1103/PhysRevC.93.011901} {\bibfield  {journal} {\bibinfo
  {journal} {Phys. Rev. C}\ }\textbf {\bibinfo {volume} {93}},\ \bibinfo
  {pages} {011901} (\bibinfo {year} {2016})},\ \Eprint
  {http://arxiv.org/abs/1506.07834} {arXiv:1506.07834 [nucl-ex]} \BibitemShut
  {NoStop}%
\bibitem [{\citenamefont {Adam}\ \emph {et~al.}(2019)\citenamefont {Adam} \emph
  {et~al.}}]{Adam:2019xmk}%
  \BibitemOpen
  \bibfield  {author} {\bibinfo {author} {\bibfnamefont {J.}~\bibnamefont
  {Adam}} \emph {et~al.} (\bibinfo {collaboration} {STAR}),\ }\href {\doibase
  10.1103/PhysRevC.100.014902} {\bibfield  {journal} {\bibinfo  {journal}
  {Phys. Rev.}\ }\textbf {\bibinfo {volume} {C100}},\ \bibinfo {pages} {014902}
  (\bibinfo {year} {2019})},\ \Eprint {http://arxiv.org/abs/1903.05370}
  {arXiv:1903.05370 [nucl-ex]} \BibitemShut {NoStop}%
\bibitem [{\citenamefont {Adamczewski-Musch}\ \emph {et~al.}(2020)\citenamefont
  {Adamczewski-Musch} \emph {et~al.}}]{Adamczewski-Musch:2020slf}%
  \BibitemOpen
  \bibfield  {author} {\bibinfo {author} {\bibfnamefont {J.}~\bibnamefont
  {Adamczewski-Musch}} \emph {et~al.} (\bibinfo {collaboration} {HADES}),\
  }\href {\doibase 10.1103/PhysRevC.102.024914} {\bibfield  {journal} {\bibinfo
   {journal} {Phys. Rev. C}\ }\textbf {\bibinfo {volume} {102}},\ \bibinfo
  {pages} {024914} (\bibinfo {year} {2020})},\ \Eprint
  {http://arxiv.org/abs/2002.08701} {arXiv:2002.08701 [nucl-ex]} \BibitemShut
  {NoStop}%
\bibitem [{\citenamefont {Adam}\ \emph {et~al.}(2020)\citenamefont {Adam} \emph
  {et~al.}}]{Adam:2020unf}%
  \BibitemOpen
  \bibfield  {author} {\bibinfo {author} {\bibfnamefont {J.}~\bibnamefont
  {Adam}} \emph {et~al.} (\bibinfo {collaboration} {STAR}),\ }\href@noop {} {\
  (\bibinfo {year} {2020})},\ \Eprint {http://arxiv.org/abs/2001.02852}
  {arXiv:2001.02852 [nucl-ex]} \BibitemShut {NoStop}%
\bibitem [{\citenamefont {Luo}\ \emph {et~al.}(2013)\citenamefont {Luo},
  \citenamefont {Xu}, \citenamefont {Mohanty},\ and\ \citenamefont
  {Xu}}]{Luo:2013bmi}%
  \BibitemOpen
  \bibfield  {author} {\bibinfo {author} {\bibfnamefont {X.}~\bibnamefont
  {Luo}}, \bibinfo {author} {\bibfnamefont {J.}~\bibnamefont {Xu}}, \bibinfo
  {author} {\bibfnamefont {B.}~\bibnamefont {Mohanty}}, \ and\ \bibinfo
  {author} {\bibfnamefont {N.}~\bibnamefont {Xu}},\ }\href {\doibase
  10.1088/0954-3899/40/10/105104} {\bibfield  {journal} {\bibinfo  {journal}
  {J. Phys.}\ }\textbf {\bibinfo {volume} {G40}},\ \bibinfo {pages} {105104}
  (\bibinfo {year} {2013})},\ \Eprint {http://arxiv.org/abs/1302.2332}
  {arXiv:1302.2332 [nucl-ex]} \BibitemShut {NoStop}%
\bibitem [{\citenamefont {Xu}\ \emph {et~al.}(2016)\citenamefont {Xu},
  \citenamefont {Yu}, \citenamefont {Liu},\ and\ \citenamefont
  {Luo}}]{Xu:2016qjd}%
  \BibitemOpen
  \bibfield  {author} {\bibinfo {author} {\bibfnamefont {J.}~\bibnamefont
  {Xu}}, \bibinfo {author} {\bibfnamefont {S.}~\bibnamefont {Yu}}, \bibinfo
  {author} {\bibfnamefont {F.}~\bibnamefont {Liu}}, \ and\ \bibinfo {author}
  {\bibfnamefont {X.}~\bibnamefont {Luo}},\ }\href {\doibase
  10.1103/PhysRevC.94.024901} {\bibfield  {journal} {\bibinfo  {journal} {Phys.
  Rev. C}\ }\textbf {\bibinfo {volume} {94}},\ \bibinfo {pages} {024901}
  (\bibinfo {year} {2016})},\ \Eprint {http://arxiv.org/abs/1606.03900}
  {arXiv:1606.03900 [nucl-ex]} \BibitemShut {NoStop}%
\bibitem [{\citenamefont {Zhou}\ \emph {et~al.}(2017)\citenamefont {Zhou},
  \citenamefont {Xu}, \citenamefont {Luo},\ and\ \citenamefont
  {Liu}}]{Zhou:2017jfk}%
  \BibitemOpen
  \bibfield  {author} {\bibinfo {author} {\bibfnamefont {C.}~\bibnamefont
  {Zhou}}, \bibinfo {author} {\bibfnamefont {J.}~\bibnamefont {Xu}}, \bibinfo
  {author} {\bibfnamefont {X.}~\bibnamefont {Luo}}, \ and\ \bibinfo {author}
  {\bibfnamefont {F.}~\bibnamefont {Liu}},\ }\href {\doibase
  10.1103/PhysRevC.96.014909} {\bibfield  {journal} {\bibinfo  {journal} {Phys.
  Rev.}\ }\textbf {\bibinfo {volume} {C96}},\ \bibinfo {pages} {014909}
  (\bibinfo {year} {2017})},\ \Eprint {http://arxiv.org/abs/1703.09114}
  {arXiv:1703.09114 [nucl-ex]} \BibitemShut {NoStop}%
\bibitem [{\citenamefont {Chatterjee}\ \emph {et~al.}(2020)\citenamefont
  {Chatterjee}, \citenamefont {Zhang}, \citenamefont {Zeng}, \citenamefont
  {Sahoo},\ and\ \citenamefont {Luo}}]{Chatterjee:2019fey}%
  \BibitemOpen
  \bibfield  {author} {\bibinfo {author} {\bibfnamefont {A.}~\bibnamefont
  {Chatterjee}}, \bibinfo {author} {\bibfnamefont {Y.}~\bibnamefont {Zhang}},
  \bibinfo {author} {\bibfnamefont {J.}~\bibnamefont {Zeng}}, \bibinfo {author}
  {\bibfnamefont {N.~R.}\ \bibnamefont {Sahoo}}, \ and\ \bibinfo {author}
  {\bibfnamefont {X.}~\bibnamefont {Luo}},\ }\href {\doibase
  10.1103/PhysRevC.101.034902} {\bibfield  {journal} {\bibinfo  {journal}
  {Phys. Rev. C}\ }\textbf {\bibinfo {volume} {101}},\ \bibinfo {pages}
  {034902} (\bibinfo {year} {2020})},\ \Eprint
  {http://arxiv.org/abs/1910.08004} {arXiv:1910.08004 [nucl-ex]} \BibitemShut
  {NoStop}%
\bibitem [{\citenamefont {Chatterjee}\ \emph {et~al.}(2016)\citenamefont
  {Chatterjee}, \citenamefont {Chatterjee}, \citenamefont {Nayak},\ and\
  \citenamefont {Sahoo}}]{Chatterjee:2016mve}%
  \BibitemOpen
  \bibfield  {author} {\bibinfo {author} {\bibfnamefont {A.}~\bibnamefont
  {Chatterjee}}, \bibinfo {author} {\bibfnamefont {S.}~\bibnamefont
  {Chatterjee}}, \bibinfo {author} {\bibfnamefont {T.~K.}\ \bibnamefont
  {Nayak}}, \ and\ \bibinfo {author} {\bibfnamefont {N.~R.}\ \bibnamefont
  {Sahoo}},\ }\href {\doibase 10.1088/0954-3899/43/12/125103} {\bibfield
  {journal} {\bibinfo  {journal} {J. Phys.}\ }\textbf {\bibinfo {volume}
  {G43}},\ \bibinfo {pages} {125103} (\bibinfo {year} {2016})},\ \Eprint
  {http://arxiv.org/abs/1606.09573} {arXiv:1606.09573 [nucl-ex]} \BibitemShut
  {NoStop}%
\bibitem [{\citenamefont {Ye}\ \emph {et~al.}(2018)\citenamefont {Ye},
  \citenamefont {Wang}, \citenamefont {Steinheimer}, \citenamefont {Nara},
  \citenamefont {Xu}, \citenamefont {Li}, \citenamefont {Lu}, \citenamefont
  {Li},\ and\ \citenamefont {Stoecker}}]{Ye:2018vbc}%
  \BibitemOpen
  \bibfield  {author} {\bibinfo {author} {\bibfnamefont {Y.}~\bibnamefont
  {Ye}}, \bibinfo {author} {\bibfnamefont {Y.}~\bibnamefont {Wang}}, \bibinfo
  {author} {\bibfnamefont {J.}~\bibnamefont {Steinheimer}}, \bibinfo {author}
  {\bibfnamefont {Y.}~\bibnamefont {Nara}}, \bibinfo {author} {\bibfnamefont
  {H.-j.}\ \bibnamefont {Xu}}, \bibinfo {author} {\bibfnamefont
  {P.}~\bibnamefont {Li}}, \bibinfo {author} {\bibfnamefont {D.}~\bibnamefont
  {Lu}}, \bibinfo {author} {\bibfnamefont {Q.}~\bibnamefont {Li}}, \ and\
  \bibinfo {author} {\bibfnamefont {H.}~\bibnamefont {Stoecker}},\ }\href
  {\doibase 10.1103/PhysRevC.98.054620} {\bibfield  {journal} {\bibinfo
  {journal} {Phys. Rev. C}\ }\textbf {\bibinfo {volume} {98}},\ \bibinfo
  {pages} {054620} (\bibinfo {year} {2018})},\ \Eprint
  {http://arxiv.org/abs/1808.06342} {arXiv:1808.06342 [nucl-th]} \BibitemShut
  {NoStop}%
\bibitem [{\citenamefont {Zhang}\ \emph {et~al.}(2020)\citenamefont {Zhang},
  \citenamefont {He}, \citenamefont {Liu}, \citenamefont {Yang},\ and\
  \citenamefont {Luo}}]{Zhang:2019lqz}%
  \BibitemOpen
  \bibfield  {author} {\bibinfo {author} {\bibfnamefont {Y.}~\bibnamefont
  {Zhang}}, \bibinfo {author} {\bibfnamefont {S.}~\bibnamefont {He}}, \bibinfo
  {author} {\bibfnamefont {H.}~\bibnamefont {Liu}}, \bibinfo {author}
  {\bibfnamefont {Z.}~\bibnamefont {Yang}}, \ and\ \bibinfo {author}
  {\bibfnamefont {X.}~\bibnamefont {Luo}},\ }\href {\doibase
  10.1103/PhysRevC.101.034909} {\bibfield  {journal} {\bibinfo  {journal}
  {Phys. Rev. C}\ }\textbf {\bibinfo {volume} {101}},\ \bibinfo {pages}
  {034909} (\bibinfo {year} {2020})},\ \Eprint
  {http://arxiv.org/abs/1905.01095} {arXiv:1905.01095 [nucl-ex]} \BibitemShut
  {NoStop}%
\bibitem [{\citenamefont {Westfall}(2015)}]{Westfall:2014fwa}%
  \BibitemOpen
  \bibfield  {author} {\bibinfo {author} {\bibfnamefont {G.~D.}\ \bibnamefont
  {Westfall}},\ }\href {\doibase 10.1103/PhysRevC.92.024902} {\bibfield
  {journal} {\bibinfo  {journal} {Phys. Rev. C}\ }\textbf {\bibinfo {volume}
  {92}},\ \bibinfo {pages} {024902} (\bibinfo {year} {2015})},\ \Eprint
  {http://arxiv.org/abs/1412.5988} {arXiv:1412.5988 [nucl-th]} \BibitemShut
  {NoStop}%
\bibitem [{\citenamefont {Zhou}\ and\ \citenamefont
  {Jia}(2018)}]{Zhou:2018fxx}%
  \BibitemOpen
  \bibfield  {author} {\bibinfo {author} {\bibfnamefont {M.}~\bibnamefont
  {Zhou}}\ and\ \bibinfo {author} {\bibfnamefont {J.}~\bibnamefont {Jia}},\
  }\href {\doibase 10.1103/PhysRevC.98.044903} {\bibfield  {journal} {\bibinfo
  {journal} {Phys. Rev. C}\ }\textbf {\bibinfo {volume} {98}},\ \bibinfo
  {pages} {044903} (\bibinfo {year} {2018})},\ \Eprint
  {http://arxiv.org/abs/1803.01812} {arXiv:1803.01812 [nucl-th]} \BibitemShut
  {NoStop}%
\bibitem [{\citenamefont {Nara}\ \emph {et~al.}(2000)\citenamefont {Nara},
  \citenamefont {Otuka}, \citenamefont {Ohnishi}, \citenamefont {Niita},\ and\
  \citenamefont {Chiba}}]{Nara:1999dz}%
  \BibitemOpen
  \bibfield  {author} {\bibinfo {author} {\bibfnamefont {Y.}~\bibnamefont
  {Nara}}, \bibinfo {author} {\bibfnamefont {N.}~\bibnamefont {Otuka}},
  \bibinfo {author} {\bibfnamefont {A.}~\bibnamefont {Ohnishi}}, \bibinfo
  {author} {\bibfnamefont {K.}~\bibnamefont {Niita}}, \ and\ \bibinfo {author}
  {\bibfnamefont {S.}~\bibnamefont {Chiba}},\ }\href {\doibase
  10.1103/PhysRevC.61.024901} {\bibfield  {journal} {\bibinfo  {journal} {Phys.
  Rev. C}\ }\textbf {\bibinfo {volume} {61}},\ \bibinfo {pages} {024901}
  (\bibinfo {year} {2000})},\ \Eprint {http://arxiv.org/abs/nucl-th/9904059}
  {arXiv:nucl-th/9904059} \BibitemShut {NoStop}%
\bibitem [{\citenamefont {Nara}(2019)}]{nara2019jam}%
  \BibitemOpen
  \bibfield  {author} {\bibinfo {author} {\bibfnamefont {Y.}~\bibnamefont
  {Nara}},\ }in\ \href@noop {} {\emph {\bibinfo {booktitle} {EPJ Web of
  Conferences}}},\ Vol.\ \bibinfo {volume} {208}\ (\bibinfo {organization} {EDP
  Sciences},\ \bibinfo {year} {2019})\ p.\ \bibinfo {pages} {11004}\BibitemShut
  {NoStop}%
\bibitem [{\citenamefont {Isse}\ \emph {et~al.}(2005)\citenamefont {Isse},
  \citenamefont {Ohnishi}, \citenamefont {Otuka}, \citenamefont {Sahu},\ and\
  \citenamefont {Nara}}]{Isse:2005nk}%
  \BibitemOpen
  \bibfield  {author} {\bibinfo {author} {\bibfnamefont {M.}~\bibnamefont
  {Isse}}, \bibinfo {author} {\bibfnamefont {A.}~\bibnamefont {Ohnishi}},
  \bibinfo {author} {\bibfnamefont {N.}~\bibnamefont {Otuka}}, \bibinfo
  {author} {\bibfnamefont {P.}~\bibnamefont {Sahu}}, \ and\ \bibinfo {author}
  {\bibfnamefont {Y.}~\bibnamefont {Nara}},\ }\href {\doibase
  10.1103/PhysRevC.72.064908} {\bibfield  {journal} {\bibinfo  {journal} {Phys.
  Rev. C}\ }\textbf {\bibinfo {volume} {72}},\ \bibinfo {pages} {064908}
  (\bibinfo {year} {2005})},\ \Eprint {http://arxiv.org/abs/nucl-th/0502058}
  {arXiv:nucl-th/0502058} \BibitemShut {NoStop}%
\bibitem [{\citenamefont {He}\ \emph {et~al.}(2016)\citenamefont {He},
  \citenamefont {Luo}, \citenamefont {Nara}, \citenamefont {Esumi},\ and\
  \citenamefont {Xu}}]{he2016effects}%
  \BibitemOpen
  \bibfield  {author} {\bibinfo {author} {\bibfnamefont {S.}~\bibnamefont
  {He}}, \bibinfo {author} {\bibfnamefont {X.}~\bibnamefont {Luo}}, \bibinfo
  {author} {\bibfnamefont {Y.}~\bibnamefont {Nara}}, \bibinfo {author}
  {\bibfnamefont {S.}~\bibnamefont {Esumi}}, \ and\ \bibinfo {author}
  {\bibfnamefont {N.}~\bibnamefont {Xu}},\ }\href@noop {} {\bibfield  {journal}
  {\bibinfo  {journal} {Physics Letters B}\ }\textbf {\bibinfo {volume}
  {762}},\ \bibinfo {pages} {296} (\bibinfo {year} {2016})}\BibitemShut
  {NoStop}%
\bibitem [{\citenamefont {Nara}\ \emph {et~al.}(2016)\citenamefont {Nara},
  \citenamefont {Niemi}, \citenamefont {Ohnishi},\ and\ \citenamefont
  {St{\"o}cker}}]{Nara:2016phs}%
  \BibitemOpen
  \bibfield  {author} {\bibinfo {author} {\bibfnamefont {Y.}~\bibnamefont
  {Nara}}, \bibinfo {author} {\bibfnamefont {H.}~\bibnamefont {Niemi}},
  \bibinfo {author} {\bibfnamefont {A.}~\bibnamefont {Ohnishi}}, \ and\
  \bibinfo {author} {\bibfnamefont {H.}~\bibnamefont {St{\"o}cker}},\ }\href
  {\doibase 10.1103/PhysRevC.94.034906} {\bibfield  {journal} {\bibinfo
  {journal} {Phys. Rev. C}\ }\textbf {\bibinfo {volume} {94}},\ \bibinfo
  {pages} {034906} (\bibinfo {year} {2016})},\ \Eprint
  {http://arxiv.org/abs/1601.07692} {arXiv:1601.07692 [hep-ph]} \BibitemShut
  {NoStop}%
\bibitem [{\citenamefont {Liu}\ \emph {et~al.}(2020)\citenamefont {Liu},
  \citenamefont {Zhang}, \citenamefont {He}, \citenamefont {Sun}, \citenamefont
  {Yu},\ and\ \citenamefont {Luo}}]{Liu:2019nii}%
  \BibitemOpen
  \bibfield  {author} {\bibinfo {author} {\bibfnamefont {H.}~\bibnamefont
  {Liu}}, \bibinfo {author} {\bibfnamefont {D.}~\bibnamefont {Zhang}}, \bibinfo
  {author} {\bibfnamefont {S.}~\bibnamefont {He}}, \bibinfo {author}
  {\bibfnamefont {K.-j.}\ \bibnamefont {Sun}}, \bibinfo {author} {\bibfnamefont
  {N.}~\bibnamefont {Yu}}, \ and\ \bibinfo {author} {\bibfnamefont
  {X.}~\bibnamefont {Luo}},\ }\href {\doibase 10.1016/j.physletb.2020.135452}
  {\bibfield  {journal} {\bibinfo  {journal} {Phys. Lett. B}\ }\textbf
  {\bibinfo {volume} {805}},\ \bibinfo {pages} {135452} (\bibinfo {year}
  {2020})},\ \Eprint {http://arxiv.org/abs/1909.09304} {arXiv:1909.09304
  [nucl-th]} \BibitemShut {NoStop}%
\bibitem [{\citenamefont {Oh}\ \emph {et~al.}(2009)\citenamefont {Oh},
  \citenamefont {Lin},\ and\ \citenamefont {Ko}}]{Oh:2009gx}%
  \BibitemOpen
  \bibfield  {author} {\bibinfo {author} {\bibfnamefont {Y.}~\bibnamefont
  {Oh}}, \bibinfo {author} {\bibfnamefont {Z.-W.}\ \bibnamefont {Lin}}, \ and\
  \bibinfo {author} {\bibfnamefont {C.~M.}\ \bibnamefont {Ko}},\ }\href
  {\doibase 10.1103/PhysRevC.80.064902} {\bibfield  {journal} {\bibinfo
  {journal} {Phys. Rev. C}\ }\textbf {\bibinfo {volume} {80}},\ \bibinfo
  {pages} {064902} (\bibinfo {year} {2009})},\ \Eprint
  {http://arxiv.org/abs/0910.1977} {arXiv:0910.1977 [nucl-th]} \BibitemShut
  {NoStop}%
\bibitem [{\citenamefont {Sombun}\ \emph {et~al.}(2019)\citenamefont {Sombun},
  \citenamefont {Tomuang}, \citenamefont {Limphirat}, \citenamefont {Hillmann},
  \citenamefont {Herold}, \citenamefont {Steinheimer}, \citenamefont {Yan},\
  and\ \citenamefont {Bleicher}}]{Sombun:2018yqh}%
  \BibitemOpen
  \bibfield  {author} {\bibinfo {author} {\bibfnamefont {S.}~\bibnamefont
  {Sombun}}, \bibinfo {author} {\bibfnamefont {K.}~\bibnamefont {Tomuang}},
  \bibinfo {author} {\bibfnamefont {A.}~\bibnamefont {Limphirat}}, \bibinfo
  {author} {\bibfnamefont {P.}~\bibnamefont {Hillmann}}, \bibinfo {author}
  {\bibfnamefont {C.}~\bibnamefont {Herold}}, \bibinfo {author} {\bibfnamefont
  {J.}~\bibnamefont {Steinheimer}}, \bibinfo {author} {\bibfnamefont
  {Y.}~\bibnamefont {Yan}}, \ and\ \bibinfo {author} {\bibfnamefont
  {M.}~\bibnamefont {Bleicher}},\ }\href {\doibase 10.1103/PhysRevC.99.014901}
  {\bibfield  {journal} {\bibinfo  {journal} {Phys. Rev. C}\ }\textbf {\bibinfo
  {volume} {99}},\ \bibinfo {pages} {014901} (\bibinfo {year} {2019})},\
  \Eprint {http://arxiv.org/abs/1805.11509} {arXiv:1805.11509 [nucl-th]}
  \BibitemShut {NoStop}%
\bibitem [{\citenamefont {Deng}\ and\ \citenamefont {Ma}(2020)}]{Deng:2020zxo}%
  \BibitemOpen
  \bibfield  {author} {\bibinfo {author} {\bibfnamefont {X.}~\bibnamefont
  {Deng}}\ and\ \bibinfo {author} {\bibfnamefont {Y.}~\bibnamefont {Ma}},\
  }\href {\doibase 10.1016/j.physletb.2020.135668} {\bibfield  {journal}
  {\bibinfo  {journal} {Phys. Lett. B}\ }\textbf {\bibinfo {volume} {808}},\
  \bibinfo {pages} {135668} (\bibinfo {year} {2020})},\ \Eprint
  {http://arxiv.org/abs/2006.12337} {arXiv:2006.12337 [nucl-th]} \BibitemShut
  {NoStop}%
\bibitem [{\citenamefont {Kitazawa}\ and\ \citenamefont
  {Luo}(2017)}]{Kitazawa:2017ljq}%
  \BibitemOpen
  \bibfield  {author} {\bibinfo {author} {\bibfnamefont {M.}~\bibnamefont
  {Kitazawa}}\ and\ \bibinfo {author} {\bibfnamefont {X.}~\bibnamefont {Luo}},\
  }\href {\doibase 10.1103/PhysRevC.96.024910} {\bibfield  {journal} {\bibinfo
  {journal} {Phys. Rev. C}\ }\textbf {\bibinfo {volume} {96}},\ \bibinfo
  {pages} {024910} (\bibinfo {year} {2017})},\ \Eprint
  {http://arxiv.org/abs/1704.04909} {arXiv:1704.04909 [nucl-th]} \BibitemShut
  {NoStop}%
\bibitem [{\citenamefont {Cheng}\ \emph {et~al.}(2009)\citenamefont {Cheng}
  \emph {et~al.}}]{Cheng:2008zh}%
  \BibitemOpen
  \bibfield  {author} {\bibinfo {author} {\bibfnamefont {M.}~\bibnamefont
  {Cheng}} \emph {et~al.},\ }\href {\doibase 10.1103/PhysRevD.79.074505}
  {\bibfield  {journal} {\bibinfo  {journal} {Phys. Rev.}\ }\textbf {\bibinfo
  {volume} {D79}},\ \bibinfo {pages} {074505} (\bibinfo {year} {2009})},\
  \Eprint {http://arxiv.org/abs/0811.1006} {arXiv:0811.1006 [hep-lat]}
  \BibitemShut {NoStop}%
\bibitem [{\citenamefont {Bzdak}\ \emph {et~al.}(2017)\citenamefont {Bzdak},
  \citenamefont {Koch},\ and\ \citenamefont {Strodthoff}}]{Bzdak:2016sxg}%
  \BibitemOpen
  \bibfield  {author} {\bibinfo {author} {\bibfnamefont {A.}~\bibnamefont
  {Bzdak}}, \bibinfo {author} {\bibfnamefont {V.}~\bibnamefont {Koch}}, \ and\
  \bibinfo {author} {\bibfnamefont {N.}~\bibnamefont {Strodthoff}},\ }\href
  {\doibase 10.1103/PhysRevC.95.054906} {\bibfield  {journal} {\bibinfo
  {journal} {Phys. Rev. C}\ }\textbf {\bibinfo {volume} {95}},\ \bibinfo
  {pages} {054906} (\bibinfo {year} {2017})},\ \Eprint
  {http://arxiv.org/abs/1607.07375} {arXiv:1607.07375 [nucl-th]} \BibitemShut
  {NoStop}%
\bibitem [{\citenamefont {Ling}\ and\ \citenamefont
  {Stephanov}(2016)}]{Ling:2015yau}%
  \BibitemOpen
  \bibfield  {author} {\bibinfo {author} {\bibfnamefont {B.}~\bibnamefont
  {Ling}}\ and\ \bibinfo {author} {\bibfnamefont {M.~A.}\ \bibnamefont
  {Stephanov}},\ }\href {\doibase 10.1103/PhysRevC.93.034915} {\bibfield
  {journal} {\bibinfo  {journal} {Phys. Rev. C}\ }\textbf {\bibinfo {volume}
  {93}},\ \bibinfo {pages} {034915} (\bibinfo {year} {2016})},\ \Eprint
  {http://arxiv.org/abs/1512.09125} {arXiv:1512.09125 [nucl-th]} \BibitemShut
  {NoStop}%
\bibitem [{\citenamefont {Kendall}\ and\ \citenamefont
  {Stuart}(1943)}]{kendall1963advanced}%
  \BibitemOpen
  \bibfield  {author} {\bibinfo {author} {\bibfnamefont {M.}~\bibnamefont
  {Kendall}}\ and\ \bibinfo {author} {\bibfnamefont {A.}~\bibnamefont
  {Stuart}},\ }\href {https://books.google.com/books?id=ARrvAAAAMAAJ} {\emph
  {\bibinfo {title} {The advanced theory of statistics}}},\ \bibinfo {series}
  {The Advanced Theory of Statistics}\ No.\ \bibinfo {number} {v. 2}\ (\bibinfo
   {publisher} {Charles Griffin: London},\ \bibinfo {year} {1943})\BibitemShut
  {NoStop}%
\bibitem [{\citenamefont {Luo}(2012)}]{Luo:2011tp}%
  \BibitemOpen
  \bibfield  {author} {\bibinfo {author} {\bibfnamefont {X.}~\bibnamefont
  {Luo}},\ }\href {\doibase 10.1088/0954-3899/39/2/025008} {\bibfield
  {journal} {\bibinfo  {journal} {J. Phys.}\ }\textbf {\bibinfo {volume}
  {G39}},\ \bibinfo {pages} {025008} (\bibinfo {year} {2012})},\ \Eprint
  {http://arxiv.org/abs/1109.0593} {arXiv:1109.0593 [physics.data-an]}
  \BibitemShut {NoStop}%
\bibitem [{\citenamefont {Luo}(2015)}]{Luo:2014rea}%
  \BibitemOpen
  \bibfield  {author} {\bibinfo {author} {\bibfnamefont {X.}~\bibnamefont
  {Luo}},\ }\href {\doibase 10.1103/PhysRevC.94.059901} {\bibfield  {journal}
  {\bibinfo  {journal} {Phys. Rev. C}\ }\textbf {\bibinfo {volume} {91}},\
  \bibinfo {pages} {034907} (\bibinfo {year} {2015})},\ \Eprint
  {http://arxiv.org/abs/1410.3914} {arXiv:1410.3914 [physics.data-an]}
  \BibitemShut {NoStop}%
\bibitem [{\citenamefont {Butler}\ and\ \citenamefont
  {Pearson}(1963)}]{Butler:1963pp}%
  \BibitemOpen
  \bibfield  {author} {\bibinfo {author} {\bibfnamefont {S.}~\bibnamefont
  {Butler}}\ and\ \bibinfo {author} {\bibfnamefont {C.}~\bibnamefont
  {Pearson}},\ }\href {\doibase 10.1103/PhysRev.129.836} {\bibfield  {journal}
  {\bibinfo  {journal} {Phys. Rev.}\ }\textbf {\bibinfo {volume} {129}},\
  \bibinfo {pages} {836} (\bibinfo {year} {1963})}\BibitemShut {NoStop}%
\bibitem [{\citenamefont {Nagle}\ \emph {et~al.}(1996)\citenamefont {Nagle},
  \citenamefont {Kumar}, \citenamefont {Kusnezov}, \citenamefont {Sorge},\ and\
  \citenamefont {Mattiello}}]{nagle1996coalescence}%
  \BibitemOpen
  \bibfield  {author} {\bibinfo {author} {\bibfnamefont {J.}~\bibnamefont
  {Nagle}}, \bibinfo {author} {\bibfnamefont {B.}~\bibnamefont {Kumar}},
  \bibinfo {author} {\bibfnamefont {D.}~\bibnamefont {Kusnezov}}, \bibinfo
  {author} {\bibfnamefont {H.}~\bibnamefont {Sorge}}, \ and\ \bibinfo {author}
  {\bibfnamefont {R.}~\bibnamefont {Mattiello}},\ }\href@noop {} {\bibfield
  {journal} {\bibinfo  {journal} {Physical Review C}\ }\textbf {\bibinfo
  {volume} {53}},\ \bibinfo {pages} {367} (\bibinfo {year} {1996})}\BibitemShut
  {NoStop}%
\bibitem [{\citenamefont {Feckov{\'a}}\ \emph {et~al.}(2015)\citenamefont
  {Feckov{\'a}}, \citenamefont {Steinheimer}, \citenamefont {Tom{\'a}\v~sik},\
  and\ \citenamefont {Bleicher}}]{Feckova:2015qza}%
  \BibitemOpen
  \bibfield  {author} {\bibinfo {author} {\bibfnamefont {Z.}~\bibnamefont
  {Feckov{\'a}}}, \bibinfo {author} {\bibfnamefont {J.}~\bibnamefont
  {Steinheimer}}, \bibinfo {author} {\bibfnamefont {B.}~\bibnamefont
  {Tom{\'a}\v~sik}}, \ and\ \bibinfo {author} {\bibfnamefont {M.}~\bibnamefont
  {Bleicher}},\ }\href {\doibase 10.1103/PhysRevC.92.064908} {\bibfield
  {journal} {\bibinfo  {journal} {Phys. Rev. C}\ }\textbf {\bibinfo {volume}
  {92}},\ \bibinfo {pages} {064908} (\bibinfo {year} {2015})},\ \Eprint
  {http://arxiv.org/abs/1510.05519} {arXiv:1510.05519 [nucl-th]} \BibitemShut
  {NoStop}%
\bibitem [{\citenamefont {Li}\ \emph {et~al.}(2020)\citenamefont {Li},
  \citenamefont {Wang}, \citenamefont {L{\"u}}, \citenamefont {Li},
  \citenamefont {Li},\ and\ \citenamefont {Liu}}]{Li:2020qqn}%
  \BibitemOpen
  \bibfield  {author} {\bibinfo {author} {\bibfnamefont {F.}~\bibnamefont
  {Li}}, \bibinfo {author} {\bibfnamefont {Y.}~\bibnamefont {Wang}}, \bibinfo
  {author} {\bibfnamefont {H.}~\bibnamefont {L{\"u}}}, \bibinfo {author}
  {\bibfnamefont {P.}~\bibnamefont {Li}}, \bibinfo {author} {\bibfnamefont
  {Q.}~\bibnamefont {Li}}, \ and\ \bibinfo {author} {\bibfnamefont
  {F.}~\bibnamefont {Liu}},\ }\href@noop {} {\  (\bibinfo {year} {2020})},\
  \Eprint {http://arxiv.org/abs/2008.11540} {arXiv:2008.11540 [nucl-th]}
  \BibitemShut {NoStop}%
\bibitem [{\citenamefont {Kuttan}\ \emph {et~al.}(2020)\citenamefont {Kuttan},
  \citenamefont {Steinheimer}, \citenamefont {Zhou}, \citenamefont
  {Redelbach},\ and\ \citenamefont {Stoecker}}]{Kuttan:2020kha}%
  \BibitemOpen
  \bibfield  {author} {\bibinfo {author} {\bibfnamefont {M.~O.}\ \bibnamefont
  {Kuttan}}, \bibinfo {author} {\bibfnamefont {J.}~\bibnamefont {Steinheimer}},
  \bibinfo {author} {\bibfnamefont {K.}~\bibnamefont {Zhou}}, \bibinfo {author}
  {\bibfnamefont {A.}~\bibnamefont {Redelbach}}, \ and\ \bibinfo {author}
  {\bibfnamefont {H.}~\bibnamefont {Stoecker}},\ }\href@noop {} {\  (\bibinfo
  {year} {2020})},\ \Eprint {http://arxiv.org/abs/2009.01584} {arXiv:2009.01584
  [hep-ph]} \BibitemShut {NoStop}%
\bibitem [{\citenamefont {He}\ and\ \citenamefont {Luo}(2017)}]{He:2017zpg}%
  \BibitemOpen
  \bibfield  {author} {\bibinfo {author} {\bibfnamefont {S.}~\bibnamefont
  {He}}\ and\ \bibinfo {author} {\bibfnamefont {X.}~\bibnamefont {Luo}},\
  }\href {\doibase 10.1016/j.physletb.2017.10.030} {\bibfield  {journal}
  {\bibinfo  {journal} {Phys. Lett.}\ }\textbf {\bibinfo {volume} {B774}},\
  \bibinfo {pages} {623} (\bibinfo {year} {2017})},\ \Eprint
  {http://arxiv.org/abs/1704.00423} {arXiv:1704.00423 [nucl-ex]} \BibitemShut
  {NoStop}%
\bibitem [{\citenamefont {Poberezhnyuk}\ \emph {et~al.}(2020)\citenamefont
  {Poberezhnyuk}, \citenamefont {Savchuk}, \citenamefont {Gorenstein},
  \citenamefont {Vovchenko}, \citenamefont {Taradiy}, \citenamefont {Begun},
  \citenamefont {Satarov}, \citenamefont {Steinheimer},\ and\ \citenamefont
  {Stoecker}}]{Poberezhnyuk:2020ayn}%
  \BibitemOpen
  \bibfield  {author} {\bibinfo {author} {\bibfnamefont {R.~V.}\ \bibnamefont
  {Poberezhnyuk}}, \bibinfo {author} {\bibfnamefont {O.}~\bibnamefont
  {Savchuk}}, \bibinfo {author} {\bibfnamefont {M.~I.}\ \bibnamefont
  {Gorenstein}}, \bibinfo {author} {\bibfnamefont {V.}~\bibnamefont
  {Vovchenko}}, \bibinfo {author} {\bibfnamefont {K.}~\bibnamefont {Taradiy}},
  \bibinfo {author} {\bibfnamefont {V.~V.}\ \bibnamefont {Begun}}, \bibinfo
  {author} {\bibfnamefont {L.}~\bibnamefont {Satarov}}, \bibinfo {author}
  {\bibfnamefont {J.}~\bibnamefont {Steinheimer}}, \ and\ \bibinfo {author}
  {\bibfnamefont {H.}~\bibnamefont {Stoecker}},\ }\href {\doibase
  10.1103/PhysRevC.102.024908} {\bibfield  {journal} {\bibinfo  {journal}
  {Phys. Rev. C}\ }\textbf {\bibinfo {volume} {102}},\ \bibinfo {pages}
  {024908} (\bibinfo {year} {2020})},\ \Eprint
  {http://arxiv.org/abs/2004.14358} {arXiv:2004.14358 [hep-ph]} \BibitemShut
  {NoStop}%
\bibitem [{\citenamefont {Braun-Munzinger}\ \emph {et~al.}(2020)\citenamefont
  {Braun-Munzinger}, \citenamefont {Friman}, \citenamefont {Redlich},
  \citenamefont {Rustamov},\ and\ \citenamefont
  {Stachel}}]{Braun-Munzinger:2020jbk}%
  \BibitemOpen
  \bibfield  {author} {\bibinfo {author} {\bibfnamefont {P.}~\bibnamefont
  {Braun-Munzinger}}, \bibinfo {author} {\bibfnamefont {B.}~\bibnamefont
  {Friman}}, \bibinfo {author} {\bibfnamefont {K.}~\bibnamefont {Redlich}},
  \bibinfo {author} {\bibfnamefont {A.}~\bibnamefont {Rustamov}}, \ and\
  \bibinfo {author} {\bibfnamefont {J.}~\bibnamefont {Stachel}},\ }\href@noop
  {} {\  (\bibinfo {year} {2020})},\ \Eprint {http://arxiv.org/abs/2007.02463}
  {arXiv:2007.02463 [nucl-th]} \BibitemShut {NoStop}%
\bibitem [{\citenamefont {Pratt}\ and\ \citenamefont
  {Steinhorst}(2020)}]{Pratt:2020ekp}%
  \BibitemOpen
  \bibfield  {author} {\bibinfo {author} {\bibfnamefont {S.}~\bibnamefont
  {Pratt}}\ and\ \bibinfo {author} {\bibfnamefont {R.}~\bibnamefont
  {Steinhorst}},\ }\href@noop {} {\  (\bibinfo {year} {2020})},\ \Eprint
  {http://arxiv.org/abs/2008.08623} {arXiv:2008.08623 [nucl-th]} \BibitemShut
  {NoStop}%
\end{thebibliography}%

\end{document}